\documentstyle[12pt,fleqn,rotate,epsfig]{article}
\newcommand{\be}{\begin{equation}}
\newcommand{\ee}{\end{equation}}
\newcommand{\bea}{\begin{eqnarray}}
\newcommand{\eea}{\end{eqnarray}}

\begin{document}
\title{Semiclassical transport of particles with dynamical
spectral functions \footnote{supported by GSI Darmstadt}}
\author{W. Cassing and S. Juchem\\
Institut f\"ur Theoretische Physik, Universit\"at Giessen\\
35392 Giessen, Germany}
\maketitle
\begin{abstract}
The conventional transport of particles in the on-shell
quasiparticle limit is extended to particles of finite life time
by means of a spectral function $A(X,\vec{P},M^2)$ for a particle
moving in an area of complex self-energy $\Sigma^{ret}_{X} = Re
\Sigma^{ret}_{X} -i \Gamma_{X}/2$. Starting from the Kadanoff-Baym
equations we derive in first order gradient expansion equations of
motion for testparticles with respect to their time evolution in
$\vec{X}, \vec{P}$ and $M^2$. The off-shell propagation is
demonstrated for a couple of model cases that simulate
hadron-nucleus collisions. In case of nucleus-nucleus collisions
the imaginary part of the hadron self-energy $\Gamma_X$ is
determined by the local space-time dependent collision rate
dynamically. A first application is presented for $A + A$
reactions up to 95 A MeV, where the effects from the off-shell
propagation of nucleons are discussed with respect to high energy
proton spectra, high energy photon production as well as kaon
yields in comparison to the available data from GANIL.
\end{abstract}

\vspace{2cm}

\newpage
\section{Introduction}
\label{introduction} The many-body theory of strongly interacting
particles out of equilibrium is a challenging problem since a
couple of decades. Many approaches based on the Martin-Schwinger
hierarchy of Green functions \cite{js61} have been formulated
\cite{Wang1,CaWa,cs85,md90,mh94} and applied to model
cases. Nowadays, the dynamical description of strongly interacting
systems out of equilibrium is dominantly based on transport
theories and efficient numerical recipies have been set up for the
solution of the coupled channel transport equations
\cite{Stoecker,Bertsch,CMMN,Cass,T3,URQMD,CB99} (and Refs.
therein). These transport approaches have been derived either from
the Kadanoff-Baym equations \cite{kb62} in Refs.
\cite{pd841,Bot,Mal,ph95,gl98} or from the hierarchy of connected
equal-time Green functions \cite{Wang1,Zuo} in Refs.
\cite{CaWa,Cass,CNW} by applying a Wigner transformation and
restricting to first order in the derivatives of the phase-space
variables ($X, P$).

However, as recognized early in these derivations \cite{CaWa,Mal},
the on-shell quasiparticle limit, that invoked additionally a
reduction of the $8N$-dimensional phase-space to $7N$ independent
degrees of freedom, where $N$ denotes the number of particles in
the system, should not be adequate for particles of short life
time and/or high collision rates. Therefore, transport
formulations for quasiparticles with dynamical spectral functions
have been presented in the past \cite{ph95,Fauser} providing a
formal basis for an extension of the presently applied transport
models
\cite{Bertsch,CMMN,URQMD,Koreview,Aich,T2,LV,RQMD,ART,Kah2,Ehehalt}.
Another branch of extensions has been developed to include
stochastic fluctuations in the collision terms
\cite{Ayik,Randrup,Burgio,BL}; these models will be denoted as
Boltzmann-Langevin (BL) approaches in spite of the different
numerical approximation schemes. Apart from the transport
extensions mentioned above there is a further branch discussing
the effects from nonlocal collisions terms
\cite{Rudy1,Rudy2,Pavel,Mora1,Mora2,Mora3}. In this work, however,
we will restrict to local collision terms (for simplicity) and
concentrate on the propagation of dynamical spectral functions.
Recipies to include nonlocal effects in collision terms can be
simulated in line with Ref. \cite{Mora3}.

So far the extension of transport theory to dynamical spectral functions
is limited to the formal level and only a few attempts have been made to
simulate the dynamics of broad resonances \cite{Ehe93,Effe99}.
In the nuclear physics context this is of particular importance for dilepton
studies in $\gamma, \pi, p$ + $A$ reactions as well as nucleus-nucleus
collisions since the vector mesons $\rho$ and $\omega$ are expected to
change their properties, i.e. their pole mass and width, during the
propagation through the nuclear medium \cite{Rapp,Brown91,CaRa}.
Apart from $e^+e^-$ in-medium spectroscopy the broadening of the hadron
spectral functions in the medium should also have some influence on
'subthreshold' meson production; here again the question is if an
enhanced yield might be due to a downward shift of the meson pole
mass or simply due to the broadening of its spectral function.
Up to now no consistent treatment of both, i.e. the real and imaginary
parts of the hadron self-energies, is available. In this work we
particularly address this question and develop a semiclassical transport
approach for space-time dependent complex self-energies.

The paper is organized as follows: In Section 2 we will derive the
generalized transport equations on the basis of the Kadanoff-Baym
equations \cite{kb62} following in part the formulation of Henning
\cite{ph95}. Special emphasis will be put on the difference to the
conventional approaches and to the derivation for an energy conserving
semiclassical limit. In Section 3 the resulting set of equations of
motion is solved for the case of a complex
time-dependent Woods-Saxon potential which allows to transparently
demonstrate the particle propagation in the 8-dimensional phase space
of a testparticle. A first application to nucleus-nucleus collisions
is presented in Section 4 for energies up to 95 A MeV showing in detail
the influence of nucleon off-shell propagation on the high energy proton
spectra and production of high energy $\gamma$-rays. The calculational
results here are controlled by experimental data from GANIL
\cite{exp,TAPS}. We, furthermore, calculate upper limits for kaon
production in $Ar + Ta$ reactions at 92 A MeV and compare to the
experiment from Ref. \cite{Lecolley}. A summary and discussion
of open problems concludes this study in Section 5.

\section{Derivation of semiclassical transport equations for particles
with dynamical life times}
In this Section we briefly recall the basic equations for Green functions
and particle self-energies as well as their symmetry properties that will be
exploited in the derivation of transport equations in the semiclassical limit.
\subsection{Properties of spectral functions}
\label{spf} Within the framework of the {\sl closed-time-path}
formalism \cite{js61,kb62,lk64} Green functions $S$ and
self-energies $\Sigma$ are given as path ordered quantities. They
are defined on the time contour consisting of two branches from
(+) $- \infty$ to $\infty$ and (--) from $\infty$ to $- \infty$.
For convenience these propagators and self-energies are
transformed into a $2\times 2$ matrix representation according to
their path structure \cite{cs85}, i.e. according to the time
branches (+) or (--) for $t$ and $t'$. Explicitly the Green
functions are given by
\\
\bea i \, S^{c}_{xy}  & = & i \, S^{++}_{xy}  \: = \: \; \; < \,
T^c \, \{ \Phi(x) \, \Phi^{\dagger}(y) \} \, > \, , \qquad i \,
S^{<}_{xy} \: = \: i \, S^{+-}_{xy}  \: = \: \; \; \eta < \, \{
\Phi^{\dagger}(y) \, \Phi(x) \} \, > \, , \nonumber\\ i \, S^{>}_{xy} &
= & i \, S^{-+}_{xy}  \: = \: \; \; \phantom{\pm} < \, \{ \Phi(x)
\, \Phi^{\dagger}(y) \} \, > \, , \qquad i \, S^{a}_{xy}  \: = \: i \,
S^{--}_{xy}  \: = \: \; \; < \, T^a \, \{ \Phi(x) \,
\Phi^{\dagger}(y) \} \, > . \nonumber\\ \eea\\
where the subscript $\cdot_{xy}$ denotes the dependence on the
coordinate space variables $x$ and $y$ and $T^c \, (T^a)$
represent the (anti-)time-ordering operators. In the definition of
$S^<$ the factor $\eta$ = +1 for bosons and $\eta$ = -1 for
fermions. In the following we consider for simplicity a theory for
scalar bosons. The modifications for relativistic fermions or
nonrelativistic particles will be presented in connection with the
final equations. The full Green functions are determined via the
Dyson-Schwinger equations for path-ordered quantities, here given
in $2 \times 2$ matrix representation \\
\setlength{\mathindent}{-0.5cm}
\bea
\left(
\begin{array}{cc}
S^{c}_{xy} & S^{<}_{xy} \\
S^{>}_{xy} & S^{a}_{xy}
\end{array}
\right)
\, = \,
\left(
\begin{array}{cc}
S^{c}_{0,xy} & S^{<}_{0,xy} \\
S^{>}_{0,xy} & S^{a}_{0,xy}
\end{array}
\right)
\, + \,
\left(
\begin{array}{cc}
S^{c}_{0,xz_1} & S^{<}_{0,xz_1} \\
S^{>}_{0,xz_1} & S^{a}_{0,xz_1}
\end{array}
\right)
\odot
\left(
\begin{array}{cc}
\Sigma^{c}_{z_1 z_2} & \! \! -\Sigma^{<}_{z_1 z_2} \\
\! -\Sigma^{>}_{z_1 z_2} & \! \Sigma^{a}_{z_1 z_2}
\end{array}
\right)
\odot
\left(
\begin{array}{cc}
S^{c}_{z_2 y} & \! S^{<}_{z_2 y} \\
S^{>}_{z_2 y} & \! S^{a}_{z_2 y}
\end{array}
\right)_{.}
\label{ds_matrix}
\eea\\
\setlength{\mathindent}{0.5cm}
The self-energies $\Sigma^{\cdot}$ are also defined according to
their time structure while the symbol "$\odot$" implies an
integration over the intermediate spacetime coordinates from
$-\infty$ to $\infty$. Linear combinations of diagonal and
off-diagonal matrix elements give the retarded and advanced Green
functions $S^{ret/adv}$ and self-energies $\Sigma^{ret/adv}$ \\
\bea
S^{ret}_{xy}
\: = \; S^{c}_{xy} - S^{<}_{xy}
\; = \; S^{>}_{xy} - S^{a}_{xy}  \; , \qquad \qquad
S^{adv}_{xy}
\: = \; S^{c}_{xy} - S^{>}_{xy}
\; = \; S^{<}_{xy} - S^{a}_{xy}  \; ,
\nonumber
\eea
\bea
\Sigma^{ret}_{xy}
\: = \; \Sigma^{c}_{xy} - \Sigma^{<}_{xy}
\; = \; \Sigma^{>}_{xy} - \Sigma^{a}_{xy} \; ,\quad \qquad \:
\Sigma^{adv}_{xy}
\: = \; \Sigma^{c}_{xy} - \Sigma^{>}_{xy}
\; = \; \Sigma^{<}_{xy} - \Sigma^{a}_{xy} \; .
\label{retquant}
\eea\\
Resorting equations (\ref{ds_matrix}) one obtains Dyson-Schwinger
equations for the retarded/advanced Green functions (where only
the respective self-energies are involved)
\\
\bea
\hat{S}_{0x}^{-1} \; S_{xy}^{ret,adv}
\; \: = \; \:
\delta_{xy}
\; \: + \; \:
\Sigma_{xz}^{ret,adv} \: \odot \: S_{zy}^{ret,adv} \; ,
\eea \\
and the wellknown Kadanoff-Baym equation for the Wightman function
$S^<$, \\
\bea
\hat{S}_{0x}^{-1} \; S_{xy}^{<}
\; \: = \; \:
\Sigma_{xz}^{ret} \: \odot \: S_{zy}^{<}
\; \: + \; \:
\Sigma_{xz}^{<} \: \odot \: S_{zy}^{adv} \: .
\label{kb_spatial}
\eea \\
In these equations $\hat{S}_{0x}^{-1}$ denotes the (negative)
Klein-Gordon differential operator which for bosonic field quanta
of (bare) mass $M_0$ is given by $\hat{S}_{0x}^{-1} = - (
\partial^\mu_x \partial^x_\mu + M^{2}_{0})$. The Klein-Gordon
equation is solved by the free propagators $S_0$ as\\
\bea
\hat{S}_{0x}^{-1}
\left(
\begin{array}{cc}
S_{0,xy}^c & S_{0,xy}^< \\
S_{0,xy}^> & S_{0,xy}^a
\end{array}
\right)
\: = \:
\delta_{xy}
\left(
\begin{array}{cc}
1 & 0 \\ 0 & -1
\end{array}
\right) \; , \qquad
\hat{S}_{0x}^{-1} \; S_{0,xy}^{ret,adv} \: = \: \delta_{xy} \;
\label{spatial_freeprop}
\eea\\
with the four-dimensional $\delta$-distribution $\delta_{xy} \equiv
\delta^{(4)}(x-y)$. \\ In the following one changes to the Wigner
representation via Fourier transformation of the rapidly
oscillating relative coordinate $(x-y)$ and formulates the theory
in terms of the coordinates $X = (x+y)/2$ and the momentum $P$,
\\
\bea F_{XP} \: = \: \int d^{4}(x\!-\!y) \; \; e^{i P_{\mu}
(x^{\mu}-y^{\mu})} \; \; F_{xy}. \label{wigner_transformation}
\eea\\
Since convolution integrals convert under Wigner transformations
as
\\
\bea \int d^{4}(x\!-\!y) \; \; e^{i P_{\mu} (x^{\mu}-y^{\mu})} \;
\; F_{1,xz} \odot F_{2,zy} \; = \; e^{-i \Diamond} \; F_{1,PX} \;
F_{2,PX}, \label{wigner_convolution} \eea\\
one has to deal with an infinite series in the differential
operator $\Diamond$ which is a four-dimensional generalization of
the Poisson-bracket, \\
\bea
 \Diamond \, \{ \, F_{1} \, \} \, \{ \, F_{2} \, \}
\; := \;
\frac{1}{2}
\left(
\frac{\partial F_{1}}{\partial X_{\mu}} \:
\frac{\partial F_{2}}{\partial P^{\mu}}
\; - \;
\frac{\partial F_{1}}{\partial P_{\mu}} \:
\frac{\partial F_{2}}{\partial X^{\mu}}
\right) \; .
\label{poissonoperator}
\eea\\
As a standard approximation of kinetic theory only contributions
up to first order in the gradients are considered. This is
justified if the gradients in the mean spatial coordinate $X$ are
small.
\\ Applying this approximation scheme (Wigner transformation and
neglecting all gradient terms of order $n \ge 2$) to the
Dyson-Schwinger equations of the retarded and advanced Green
functions one ends up with  \\
\bea
( \, P^2 \, - \, M_{0}^{2} \, - \, Re \Sigma^{ret}_{XP} \, ) \: Re S^{ret}_{XP}
& = & 1 \: - \:
\frac{1}{4} \, \Gamma_{XP} \: A_{XP} \; ,
\nonumber\\[0.2cm]
( \, P^2 \, - \, M_{0}^{2} \, - \, Re \Sigma^{ret}_{XP} \, ) \:
A_{XP} & = & \Gamma_{XP} \: Re S^{ret}_{XP} \; ,
\label{dsret_firstorder} \eea \\
where we have separated the retarded and advanced Green functions
as well as the self-energies into real and imaginary contributions
\bea
S_{XP}^{ret,adv}
\; = \;
Re S^{ret}_{XP}
\; \mp \;
\frac{i}{2} \, A_{XP}
\; , \qquad
\Sigma_{XP}^{ret,adv}
\; = \;
Re \Sigma^{ret}_{XP}
\; \mp \;
\frac{i}{2} \, \Gamma_{XP} \; .
\label{ret_sep}
\eea\\
The imaginary part of the retarded propagator is given (up to a
factor) by the normalized spectral function
\bea A_{XP} \: = \: i \left[ \, S_{XP}^{ret} \: - \: S_{XP}^{adv}
\, \right] \: = \: - 2 \, Im \, S^{ret}_{XP} \; , \qquad \qquad
\qquad \int \frac{d P_0^2}{4 \pi} \,  A_{XP} \; = \; 1 \; ,
\nonumber \label{spectralfunction} \eea
while the imaginary part of the self-energy is half the width
$\Gamma_{XP}$. From the algebraic equations
(\ref{dsret_firstorder}) we obtain a direct relation of the real
and the imaginary part of the propagator (provided $\Gamma_{XP} >
0$):
\bea
Re S^{ret}_{XP} \; = \;
\frac{P^{2} \: - \: M_{0}^{2} \: - \: Re \Sigma^{ret}_{XP}}{\Gamma_{XP}}
\; A_{XP} \, .
\label{dispersion}
\eea
Finally the solution for the spectral function shows a Lorentzian
shape with space-time and four-momentum dependent width
$\Gamma_{XP}$. This result is valid for bosons to the first order
in the gradient expansion,
\bea A_{XP} \; = \; \frac{ \Gamma_{XP} } {( \, P^2 \, - \,
M_{0}^{2} \, - \, Re \Sigma^{ret}_{XP} )^{2} \: + \:
\Gamma_{XP}^{2}/4} \; . \label{alg_spectral} \eea\\
For the real part of the Green function we get
\bea Re S^{ret}_{XP} \; = \; \frac{P^{2} \: - \: M_{0}^{2} \: - \:
Re \Sigma^{ret}_{XP}} {( \, P^2 \, - \, M_{0}^{2} \, - \, Re
\Sigma^{ret}_{XP} )^{2} \: + \: \Gamma_{XP}^{2}/4} \, .
\label{alg_realpart} \eea \\

\subsection{Transport equations}
In this Subsection we derive the transport equations that will be
used to describe the propagation of particles with a space-time
dependent finite life time. For this aim we start with the
Kadanoff-Baym equation (\ref{kb_spatial}) which yields in the same
approximation scheme (i.e. a first order gradient expansion of the
Wigner transformed equation) by separating real and imaginary
parts to a generalized transport equation, \\
\bea
\Diamond \, \{ \, P^{2} &-& M_{0}^{2} \: - \: Re \Sigma^{ret}_{XP} \, \}
\; \{ \, S^{<}_{XP} \, \}
\; - \;
\Diamond \, \{ \, \Sigma^{<}_{XP} \, \} \; \{ Re S^{ret}_{XP} \, \}
\nonumber\\[0.2cm]
&=&
\frac{i}{2} \:
\left[ \: \Sigma^{>}_{XP} \: S^{<}_{XP} \; - \;
          \Sigma^{<}_{XP} \: S^{>}_{XP} \: \right],
\label{general_transport}
\eea\\
and to a generalized mass-shell equation, \\
\bea
[ \, P^{2} &-& M_{0}^{2} \: - \: Re \Sigma^{ret}_{XP} \, ] \; S^{<}_{XP}
\; - \;
\Sigma^{<}_{XP} \; Re S^{ret}_{XP}
\nonumber\\[0.2cm]
&=&
\frac{1}{2} \,
\Diamond \; \{ \, \Sigma^{<}_{XP} \, \} \; \{ \, A_{XP} \, \}
\; - \;
\frac{1}{2} \,
\Diamond \; \{ \, \Gamma_{XP} \, \} \; \{ \, S^{<}_{XP} \, \} \; .
\label{general_massshell}
\eea\\
In the transport equation (\ref{general_transport}) one recognizes
on the l.h.s. the drift term  $P^{\mu} \partial_{\mu} \bullet$,
generated by the contribution $\Diamond \{ P^2 - M_0^2 \} \{
\bullet \}$, as well as the Vlasov term determined by the real
part of the retarded self-energy. On the other hand the r.h.s.
represents the collision term with its 'gain and loss' structure.
To evaluate the $\Diamond \{ \Sigma^< \} \{ Re S^{ret} \}$-term in
(\ref{general_transport}), which does not contribute in the
quasiparticle limit, it is useful to introduce distribution
functions for the Green functions and self-energies as\\
\bea \label{sep}
 i \; S_{XP}^{<} \; = \; N_{XP} \: A_{XP} \: ,
\qquad \qquad i \; S_{XP}^{>} \; = \; ( \, 1 \, + \, N_{XP} \, )
\: A_{XP} \: , \nonumber\\[0.2cm]
i \; \Sigma_{XP}^{<} \; = \; N_{XP}^{\Sigma} \: \Gamma_{XP} \: ,
\qquad \qquad
i \; \Sigma_{XP}^{>} \; = \; ( \, 1 \, + \, N_{XP}^{\Sigma} \, )
\: \Gamma_{XP} \: .
\eea\\
Following the argumentation of Botermans and Malfliet \cite{Bot}
the distribution functions $N$ and $N^{\Sigma}$ in (\ref{sep})
should be equal within the second term of the l.h.s. of
(\ref{general_transport}) within a consistent first order gradient
expansion. In order to demonstrate their argument we write
\\
\bea
\Sigma^{<}_{XP}
\; \; = \; \; - \, i \: \Gamma_{XP} \; N^{\Sigma}_{XP}
\; \; = \; \; - \, i \: \Gamma_{XP} \; N_{XP} \; \: + \; \: C_{XP} \: .
\eea\\
The 'correction' term $C_{XP}$ is proportional to the collision
term (r.h.s.) of the generalized transport equation
(\ref{general_transport}), \\
\bea C_{XP} \; = \; -\, i \: \Gamma_{XP} \; ( \, N^{\Sigma}_{XP}
\; - \; N_{XP} \, ) \; = \; i \; ( \, \Sigma^{<}_{XP} \;
S^{>}_{XP} \; - \;\Sigma^{>}_{XP} \; S^{<}_{XP} \,) \; A^{-1}_{XP}
\: ,\eea\\
which itself is of first order in the gradients. Thus, whenever a
distribution function $N^{\Sigma}$ appears within a Poisson
bracket the difference term $(N^{\Sigma} - N$) becomes of second
order in the gradients and has to be omitted for consistency. As a
consequence $N^{\Sigma}$ can be replaced by $N$ and the
self-energy $\Sigma^<$ by $S^< \cdot \Gamma / A$ in the term
$\Diamond \{\Sigma ^{<}\} \{Re S^{ret}\}$. The general transport
equation (\ref{general_transport}) then can be written as
\\
\setlength{\mathindent}{-0.5cm}
\bea A_{XP} \, \Gamma_{XP} && \!\!\!\!\! \!\!\!\!\! \left[ \,
\Diamond \; \{ \, P^2 - M_0^2 - Re \Sigma^{ret}_{XP} \, \} \; \{
\, S^<_{XP} \, \} \: - \: \frac{1}{\Gamma_{XP}} \; \Diamond
\; \{ \, \Gamma_{XP} \, \} \; \{ \, ( \, P^2 - M_0^2 - Re
\Sigma^{ret}_{XP} \, ) \, S^<_{XP} \, \} \, \right]
\nonumber\\[0.2cm] && \; = \; i \, \left[ \, \Sigma^>_{XP} \:
S^<_{XP} \: - \: \Sigma^<_{XP} \: S^>_{XP} \, \right].
\label{trans_approx} \eea \\
\setlength{\mathindent}{0.5cm}
In order to explore the physical
content of (\ref{trans_approx}) we simplify the problem by
assuming the self-energy only to depend on space-time coordinates,
i.e. $ Re \Sigma^{ret}_{XP} = Re \Sigma^{ret}_{\vec{X}},
\Gamma_{XP} = \Gamma_X$. In view of transport applications it is
now advantageous to introduce a 'virtual mass parameter' $M^{2}$
that is determined by the four-momentum of the particles and
incorporates also energy-shifts generated by the real part of the
self-energy as
\bea
M^{2} \; = \;
P^{2} \: - \: Re \Sigma_{\vec{X}}^{ret} \, .
\eea
Taking $M^{2}$ as an independent variable this fixes the energy
(for given $\vec{P}$ and $M^{2}$) to
\bea
P_{0}^{2} \; = \;
\vec{P}^{2} \: + \: M^{2} \: + \: Re \Sigma_{\vec{X}}^{ret} \, .
\label{energyfix}
\eea
In this limit the transport equation simplifies as
\bea \label{23} & \Diamond & \!\!\!\!\!\! \{ \, P^2 - M_0^2 - Re
\Sigma^{ret}_{XP} \, \} \; \{ \, S^<_{XP} \, \} \: - \:
\frac{1}{\Gamma_{XP}} \; \Diamond \; \{ \, \Gamma_{XP} \, \}
\; \{ \, ( \, P^2 - M_0^2 - Re \Sigma^{ret}_{XP} \, ) \, S^<_{XP}
\, \} \nonumber\\[0.2cm] & \longrightarrow & - P^{\mu} \,
\partial_{\mu}^X \, S^<_{XP} \: + \: \frac{1}{2} \, \left( \,
\vec{\nabla}_X \, Re \Sigma^{ret}_{\vec{X}} \, \right) \: \left(
\, \vec{\nabla}_P \, S^<_{XP} \right) \nonumber\\[0.2cm] && \: -
\: \frac{1}{2 \Gamma_X} \, \left( \, \partial_{\mu}^X \, \Gamma_X \,
\right) \: \left( \, \partial_P^{\mu} \, \left[ \, ( \, P^2 -
M_0^2 - Re \Sigma^{ret}_{\vec{X}} \, ) \; S^<_{XP} \, \right] \,
\right) \; .
\eea
To solve equation (\ref{trans_approx}) we use a testparticle
ansatz for the real quantity $F_{X,{\vec P},M^2}$, which
represents the probability to find a particle in a given
phase-space cell, \\
\bea
  F_{X\vec{P}M^2} \; = \; i \, S^{<}_{X\vec{P}M^2}
\; \sim \; \sum_{i=1}^{N} \; \delta^{(3)} ({\vec{X}} \, - \,
{\vec{X}}_i (t)) \; \; \delta^{(3)} ({\vec{P}} \, - \, {\vec{P}}_i
(t)) \; \; \delta(M^{2} \, - \, M_i^{2}(t)) \: .
\label{testparticle} \eea \\
In (\ref{testparticle}) we have formulated the testparticle ansatz in terms
of the 'mass parameter' $M^2$ instead of the energy $P_0$ to gain
the evolution in $M^2$. $F_{X{\vec P}M^2}$ then is  a solution
of (\ref{23}) if the testparticle coordinates obey the following
equations of motion \\
\bea \label{eomr} \frac{\partial {\vec X}_i}{\partial t} & = &
\frac{{\vec P}_i}{P_{0i}} \: ,
\\[0.2cm]
\label{eomp} \frac{\partial {\vec P}_i}{\partial t} & = & -
\frac{1}{2 P_{0i}} \, {\vec \nabla}_{X_i} \, Re
\Sigma^{ret}_{{\vec{X}}_i} \: - \: \frac{M_i^{2} -
M_{0}^{2}}{\Gamma_{X_i}}
\frac{1}{2 P_{0i}} \, {\vec \nabla}_{X_i}
\, \Gamma_{X_i} \: ,
\\[0.2cm]
\label{eomm1} \frac{\partial M_i^{2}}{\partial t} & = &
\frac{M_i^{2} - M_{0}^{2}}{\Gamma_{X_i}} \, \frac{d}{dt} \,
\Gamma_{X_i} \: = \frac{M_i^{2} - M_{0}^{2}}{\Gamma_{X_i}} \,
\{\frac{\partial}{\partial t} \, \Gamma_{X_i} \: + \frac{{\vec P}
}{ P_{0i}} \, {\vec \nabla}_{X_i} \, \Gamma_{X_i} \: \}. \eea \\
For $\Gamma_{X} = 0$ we obtain directly the familiar equations of
motion in the quasiparticle approximation assuming quasiparticles
states with an effective mass squared
$M_{0}^2 \, + \, Re \Sigma^{ret}_{\vec{X}}$ and a spectral function
proportional to a $\delta$-function.
This limit has been traditionally employed in
transport theories \cite{CMMN,Koreview,Aich} allthough its
applicability should be restricted to low collision rates.
Starting from the general equation (\ref{trans_approx}), which
does not invoke any on-shell approximation, the quasiparticle
limit can also be obtained for small width $\Gamma$ by integrating
out the energy $P_0$. The second term on the l.h.s. of
(\ref{general_transport}) then vanishes while all other terms are
given in terms of corresponding quasiparticle quantities (Green
functions, self energies, etc.). We note that the quasiparticle
approximation is known to be a self-consistent energy conserving
approximation. Furthermore, the conventional quasiparticle
approximation also holds for particles of finite life time if $\,
\partial_{t} \, \Gamma_{X} = \vec{\nabla} \, \Gamma_{X} = 0$, i.e.
in the low density regime with almost vanishing collision rate
$\Gamma_{coll}$ (see below). \\
 Eqs. (\ref{eomr}) - (\ref{eomm1})
fulfill energy conservation with respect to the particle
propagation if
\bea
\frac{{\vec P}}{P_{0}} \cdot {\vec \nabla}_{X}
\Gamma_{X} \: - \: \frac{d}{dt} \, \Gamma_{X} = 0 \, ,
\label{energycondition}
\eea \\
i.e. ${d M^{2}}/ {dt} = - 2 \, {\vec P} \cdot {d {\vec P}}/ {dt}$
(with respect to the additional terms involving $\Gamma_{X}$). Eq.
(\ref{energycondition}) is identically fulfilled for $\partial
\Gamma_X /\partial t = 0$ according to (\ref{eomm1}), however,
violated for $\partial \Gamma_X /\partial t \neq 0$. In view of a
semiclassical treatment we suggest to fix the energy $P_{0i}$ at
every time $t$ by adding a term $\sim \partial \Gamma_X/ \partial t$ in eq.
(\ref{eomp}), i.e.
\bea \label{eomrf} \frac{\partial {\vec X}_i}{\partial t} & = &
\frac{{\vec P}_i}{P_{0i}} \: ,
\\[0.2cm]
\label{eompf} \frac{\partial {\vec P}_i}{\partial t} & = & -
\frac{1}{2 P_{0i}} \, {\vec \nabla}_{X_i} \, Re
\Sigma^{ret}_{{\vec{X}}_i} \: - \: \frac{M_i^{2} -
M_{0}^{2}}{\Gamma_{X_i}} \left( \frac{1}{2 P_{0i}} \, {\vec
\nabla}_{X_i} \, \Gamma_{X_i} \: + \: \frac{\vec{P_i}}{2
\vec{P}^2_i} \, \frac{\partial}{\partial t} \, \Gamma_{X_i}
\right) \: ,
\\[0.2cm]
\label{eomm1f}
\frac{\partial M_i^{2}}{\partial t}
& = &
\frac{M_i^{2} - M_{0}^{2}}{\Gamma_{X_i}}
\, \frac{d}{dt} \, \Gamma_{X_i} \: .
\eea \\
Dynamical effects of the additional term in (\ref{eompf}) will be
discussed in the first applications of our model in Section 3. \\
In case of relativistic fermions with (bare) mass $m_{0}$ we have \\
\bea Re \Sigma_{X}^{ret} \; \equiv \; - V_{0}^{2} \: + \: {\vec
V}^{2} \: + \: 2 P_{0} \, V_{0} \: - \: 2 {\vec P} \cdot {\vec V}
\: + \: 2 m_{0} \, U_{S} \: + \: U_{S}^{2} \, ,
\label{relfermions_re} \eea \\
where $V_{\mu} = ( V_{0} , {\vec V})$ and $U_{S}$ denote the real
part of the vector and scalar self-energy \cite{Serot}, respectively.
Here we have neglected pseudo-scalar, pseudo-vector and tensor
contributions in the Gordon decomposition of the self-energy,
which either vanish identically or remain small in the nuclear
physics context. \\
A similar decomposition holds for $\Gamma_{X}$, i.e. \\
\bea
\Gamma_{X} \; \equiv \;
- W_{0}^{2} \: + \: {\vec W}^2
\: + \: 2 P_{0} \, W_{0} \: - \: 2 {\vec P} \cdot {\vec W}
\: + \: 2 m_{0} \, W_{S} \: + \: W_{S}^{2}
\label{rel2}
\eea \\
for the vector part ($W_{\mu}$) and scalar part ($W_{S}$) of the
imaginary self-energy.

\section{Model studies}
For our present purpose to demonstrate the physical implications of
eqs. (\ref{eomrf}) - (\ref{eomm1f}) we consider the propagation of
particles in a time-dependent complex potential of Woods-Saxon form,
i.e. \\
\setlength{\mathindent}{-0.2cm}
\bea Re \Sigma_{X}^{ret} \, - \, \frac{i}{2} \, \Gamma_{X} \; = \;
2 P_0 \left\{ \, \frac{V_{0} }{1 + \exp\{(|\vec{r}| - R)/a_0\}} \,
- \, i \left( \frac{W_{0}(t)}{1 + \exp\{(|\vec{r}| - R)/a_0\}} \:
+ \: \frac{\Gamma_V}{2} \right) \! \! \right\} \; \label{cpot} \eea \\
\setlength{\mathindent}{0.2cm}
where we have used $R=5$ fm, $a_0 = 0.6$ fm throughout the model
studies. Eqs. (\ref{eomrf}) - (\ref{eomm1f}) allow to represent
the distribution function in terms of the testparticle
distribution (\ref{testparticle}) where ${\vec r}_i(t), {\vec P}_i(t)$ and
$M^2_i(t)$ are the corresponding solutions of eqs. (\ref{eomrf}) -
(\ref{eomm1f}). We initialize all testparticles $i$ with a fixed
energy $P_{0}$ at some distance ($|{\vec{r}}(t=0)| \approx -15$
fm) on the $z$-axis with a three-momentum vector in positive
$z$-direction. The mass parameters $M_{i}(t=0)$ are selected
according to the Breit-Wigner-distribution
\bea
F(M) \; = \;
\frac{1}{2 \pi} \frac{\Gamma_V} {(M \, - \, M_0)^2
\: + \: \Gamma_V^2/4},
\label{nonrelbreitwigner}
\eea
where $\Gamma_V$ denotes the vacuum width which might be arbitrarily
small but finite (see below).
The particles are then propagated in time according to eqs. (\ref{eomrf}) -
(\ref{eomm1f}) and all one-body quantities can be evaluated from
(\ref{testparticle}).

In Fig. 1 (upper part) the results for $P_{i0}(z(t))$,
$M_{i}(z(t))$ and $P_{iz}(z(t))$ are diplayed as a function of
$z(t)$ instead of the time t. We show the evolution of 21
testparticles with mass parameters that are initially seperated by
$\Delta M = 0.05 \cdot \Gamma_{V}$ in the case of a nonvanishing
imaginary part of the potential ($W_{0}(t) = W_{0} = 70$ MeV,
$\Gamma_{V} = 0.8$ MeV) but vanishing real part of the potential
($V_{0} = 0$ MeV) (see Fig. 1 (lower part)). One recognizes that
the differences between the mass parameters increase when reaching
the potential region, which corresponds directly to a broadening
of the spectral function. The same spreading behavior is observed
for the three-momentum of the testparticles, such that the energy
$P_{0}$ is conserved throughout the whole calculation (upper
line). When leaving the potential region the splitting decreases
and the correct asymptotic solution is restored.

At next we present in Fig. 2 (upper part) a calculation
where we additionally allow for a nonvanishing real part
of the potential (i.e. $V_{0} = -20$ MeV, Fig. 2 (lower part)).
While the spreading of the mass parameter is not affected
by this change, we find a shift of the testparticle momenta
where the real part of the potential deviates from zero since here the
particles are accelerated.

While we up to now have only considered constant potentials
in time, we now introduce an explicit
time dependence corresponding to $W_{0}(t) = 100$ MeV
$(1-0.02 \,c/$fm$ \cdot t)$.
As a result the spatial reflection symmetry vanishes
for $P_{iz}(z(t))$ and $M_{i}(z(t))$ (cf. Fig. 3).
For the given time dependence the mass splitting is smaller
for given $z$ compared to a time-independent potential.
As in the former cases here also the correct free
solution is obtained for large $z$ while the energy is
strictly conserved, too.

In order to discuss the effect of the additional term $\sim
\partial \Gamma_X /\partial t$ introduced in eq. (\ref{eompf}) we
solve eqs. (\ref{eomr}) - (\ref{eomm1}) that directly stem from
the testparticle ansatz (\ref{testparticle}) for the model case presented in
Fig. 3. The corresponding results in Fig. 4 show that the time
evolution in $M^2$ is hardly effected, however, the individual
trajectories spread out in energy $P_0$ which can be traced back
to an asymptotic spread in momentum $P_z$. When integrating over
the particle spectral function it can be shown that the total
energy of the particle -- which corresponds to a quantum
mechanical state -- is conserved again due to the symmetry of the
spectral function in $(M^2-M_0^2)$ for a local width $\Gamma_X$.
Thus the original equations of motion (\ref{eomr}) - (\ref{eomm1})
conserve energy with respect to the quantum mechanical state.
However, within semiclassical transport simulations the spectral
function will not be populated symmetrically around $M_0^2$ due to
energy constraints, e.g. in reactions $pp \rightarrow pp \rho^0$
with invariant energy $\sqrt{s} \leq 2 M_p + M^\rho_0$. Thus only
the low mass fraction of the Breit-Wigner distribution will be
populated for which the energy is no longer conserved throughout
the propagation which might produce artefacts in further inelastic
reaction channels. Such artefacts do not occur in a fully quantum
mechanical theory due to the phase coherence $\sim iP_0t$ which
guarantees energy conservation for $t \rightarrow \infty$. Since
semiclassical transport simulations do not involve such phase
coherence and thus violate the uncertainty relation with respect
to energy and time, we have restored energy conservation locally
in eqs. (\ref{eomrf}) - (\ref{eomm1f}) for each testparticle
representing a tiny slice in $dM^2$ of the spectral function.

In the next example of this model study we show in Fig. 5
(upper part) the case of a broad vacuum spectral function
entering a (time-independent) nonrelativistic potential
with $V_{0} = -20$ MeV and $W_{0}(t) = W_{0} = 100$ MeV.
The vacuum width is chosen as $\Gamma_{V} = 160$ MeV, while
11 testparticle trajectories are shown with an initial separation
of the masses $\Delta M = 0.05 \cdot \Gamma_{V}$.
One observes that the spectral function is further broadened
in the complex potential zone and reaches its inital dispersion
in mass again after passing the diffractive and absorptive area.

The question remains if the testparticle distribution
(\ref{testparticle})
reproduces the local splitting in mass as expected due to quantum
mechanics, i.e. in our case a Breit-Wigner distribution
(\ref{nonrelbreitwigner}) with a local width $\Gamma_X$ = 2
$W_0(z) + \Gamma_V$. This is demonstrated in Fig. 6 where we show
the spectral function as a function of mass $M$ from the
testparticle evolution at fixed coordinate $z$ in comparison to
the quantum Breit-Wigner distribution with local width $\Gamma_X$
(full lines) for a pure imaginary potential with parameters $W_0$
= 50 MeV and vacuum width $\Gamma_V$ = 2 MeV. The differences from
the exact results in Fig. 6 are practically not visible for all
values of $z$ from - 8 fm to 8 fm. The width of the distribution
increases from 1 MeV in the vacuum ($z = \pm$ 8 fm) to 102 MeV (=
2 $W_0$+ $\Gamma_V$) in the center of the absorptive potential
($z$ = 0). Thus our off-shell quasiparticle propagation is fully
in line with the quantum mechanical result at least for
quasi-stationary quantum states.

To summarize our model results for the simple complex potential
of Woods-Saxon-type, we find that energy conservation is guaranteed
during the propagation as well as that the correct asymptotic
solutions for the spectral functions are restored. Furthermore,
in the potential region we observe a broadening of the width of the
spectral function due to the space-time dependent imaginary part
of the potential in line with  quantum mechanics.

\section{Application to nucleus-nucleus collisions}
Apart from the model studies performed in the previous Section it
is of interest, if the 'off-mass-shell approach' proposed here
leads to observable consequences in actual experiments such as for
nucleus-nucleus collisions. This implies to specify the collision
term in eq. (\ref{general_transport}). A corresponding expression
can be formulated in full analogy to Ref. \cite{CMMN} by giving
explicit approximations for $\Sigma^<$ and $\Sigma^>$ and
using detailed balance as
\be
\label{Icoll}
I_{coll}(X,\vec{P},M^2) = Tr_2 Tr_3 Tr_4 A(X,{\vec P},M^2)
A(X,{\vec P}_2, M_2 ^2) A(X,{\vec P}_3, M_3 ^2)
A(X,{\vec P}_4, M_4 ^2)
\ee
$$
\{ |T(({\vec P},M^2) + ({\vec P}_2,M_2^2)
\rightarrow ({\vec P}_3,M_3^2) + ({\vec P}_4,M_4^2))|_{{\cal A,S}}^2
\: \delta^4({P} + {P}_2 - {P}_3 - {P}_4)
$$
$$
[N_{X{\vec P}_3 M_3^2} N_{X {\vec P}_4 M_4^2} {\bar f}_{X {\vec P},M^2}
{\bar f}_{X {\vec P}_2, M_2^2}
- N_{X{\vec P} M^2} N_{X {\vec P}_2 M_2^2} {\bar f}_{X {\vec P}_3,M_3^2}
{\bar f}_{X {\vec P}_4,M_4^2} ]  \}
$$
with
\be
\label{pauli} {\bar f}_{X{\vec P}} = 1 + \eta N_{X {\vec P}} \ee
and $\eta = \pm 1$ for bosons/fermions, respectively. The indices
${\cal A,S}$ stand for the antisymmetric/symmetric matrix element
of the scattering amplitude $T$ in case of fermions/bosons. We
note that in (\ref{Icoll}) we have neglected gradient terms  which
lead to nonlocal collision terms
\cite{Rudy1,Rudy2,Mora1,Mora2,Mora3}. Their effect can
additionally be taken into account following the suggestions of
Ref. \cite{Mora3}, but are discarded here for simplicity and
transparency. \\ In eq. (\ref{Icoll}) the trace over particles
2,3,4 reads explicitly for fermions
\be
\label{trace} Tr_2 = \sum_{\sigma_2, \tau_2} \frac{1}{(2 \pi)^4}
\int d^3 P_2 \, \frac{d M^2_2}{2 \sqrt{{\vec P}_2^2 + M^2_2}}, \ee where
$\sigma_2, \tau_2$ denote the spin and isospin of particle 2. In
case of bosons we have
\be
\label{trace2} Tr_2 = \sum_{\sigma_2, \tau_2} \frac{1}{(2 \pi)^4}
\int d^3 P_2 \, \frac{d P_{0,2}^2}{2}, \ee
since here the spectral
function is normalized as
\be
\label{sb} \int \frac{d P_0^2}{4 \pi} A_B(X,P) = 1
\ee
whereas for fermions we have
\be
\label{sb1} \int \frac{d P_0}{2 \pi} A_F(X,P) =
\frac{M_0}{\sqrt{{\vec P}^2 + M^2_0}}.
\ee
It is easy to show that
the collision term (\ref{Icoll}) leads to the proper Bose or
Fermion equilibrium distributions for $t \rightarrow \infty$. \\
Neglecting the 'gain-term' in eq. (\ref{Icoll}) one recognizes
that the collisional width of the particle in the rest frame is
given by
\be
\label{gcoll} \Gamma_{coll}(X,\vec{P},M^2) = Tr_2 Tr_3 Tr_4 \;
\{|T(({\vec P},M^2) + ({\vec P}_2,M_2^2) \rightarrow ({\vec
P}_3,M_3^2) + ({\vec P}_4,M_4^2))|_{{\cal A,S}}^2 \ee
$$ A(X,{\vec P}_2,M_2^2) A(X,{\vec P}_3,M_3^2) A(X,{\vec P}_4, M_4^2)
\delta^4(P + P_2 - P_3-P_4) \ N_{X {\vec
P}_2 M_2^2} {\bar f}_{X {\vec P}_3} {\bar f}_{X {\vec P}_4} \},
$$
where as in eq. (\ref{Icoll}) local on-shell scattering processes
are assumed. We note that the extension of eq.(\ref{Icoll}) to
inelastic scattering processes (e.g. $NN \rightarrow N\Delta$) or
($\pi N \rightarrow \Delta$ ect.) is straightforward when exchanging the
elastic transition amplitude $T$ by the corresponding inelastic one and
taking care of Pauli-blocking or Bose-enhancement for the particles in the
final state. We mention that for bosons we will neglect a Bose-enhancement
since their actual phase-space density is small
for the systems of interest. \\
For particles of infinite life time in vacuum -- such as protons
-- the collisional width (\ref{gcoll}) has to be identified with
half the imaginary part of the self-energy that determines the
spectral function (\ref{nonrelbreitwigner}). Thus the transport
approach determines the particle spectral function dynamically via
(\ref{gcoll}) for all hadrons if the in-medium transition
amplitudes $T$ are known. Since in binary collisions due to energy
and momentum conservation -- once the final masses are fixed --
only the final scattering angle $\Omega = (\cos \theta, \phi)$ is
undetermined we can replace the amplitude squared in (\ref{gcoll})
as \bea \label{xsection} |T({\vec q})|_{{\cal A,S}} = \frac{4
\pi^2}{\mu^2} \ \frac{d \sigma}{d \Omega}(\sqrt{s}) \eea where
${\vec q}$ is the momentum transfered in the collision at
invariant energy $\sqrt{s}$ and $\mu$ is the reduced mass of the
scattering particles. The differential cross section $d\sigma/d
\Omega$ or $T({\vec q})$ in principle should be evaluated in the
Brueckner approach, however, in practice effective
parametrizations are employed (see below). \\
\subsection{Numerical realisation}
The following dynamical calculations are based on the conventional
HSD transport approach \cite{CB99,Ehehalt}, where for energies up
to 100 A MeV (GANIL energies) essentially the nucleon degrees of
freedom are important, since inelastic processes
$NN \rightarrow N\Delta \rightarrow \pi N, \; \pi N \rightarrow \Delta$
are suppressed. We only briefly mention that the formation cross
section in the reaction $NN \rightarrow N\Delta$ is evaluated with
the total $\Delta$-width $\Gamma_{\Delta}^{tot} = \Gamma_{\Delta}^{\pi N}
+ \Gamma_{\Delta}^{coll}$ \cite{Ehe93} and that in the decay channel
$\Delta \rightarrow \pi N$ the final nucleon state is selected by
Monte Carlo using the local spectral distribution (\ref{nonrelbreitwigner}). \\
Whereas the real part of the nucleon self-energy is
determined as in Ref. \cite{CB99} and includes an explicit momentum
dependence of the scalar and vector self-energies for nucleons in order
to qualify also for relativistic reactions,
we have to describe in more detail the implementation of the off-shell dynamics
induced by $\Gamma_{XP}= \Gamma_{coll}(X,{\vec P},M^2)$.
According to (\ref{gcoll}) the collisional width is explicitly
momentum (and energy) dependent, which introduces much larger
numerical efforts as for momentum-dependent real potentials. In view of the
limited energy range addressed here and for the purpose of
an exploratory study it is sufficient to consider a
space-time dependent collisional width $\Gamma_X$ which is
obtained by averaging (\ref{gcoll}) at each time-step
($\Delta t = 0.5 fm/c$) in each cell in coordinate space
(of size 1$fm^3$),
\be
\label{average} \Gamma_X = \frac{1}{\rho(X)} \sum_{\sigma,\tau}
\frac{1}{(2\pi)^4} \int d^3 P \frac{d M^2}{2 \sqrt{{\vec P}^2 +
M^2}} \Gamma_{X {\vec P} M^2} \ee with the local density
\be
\label{dens} \rho(X) =  \sum_{\sigma,\tau} \frac{1}{(2\pi)^4} \int
d^3 P \frac{d M^2}{2 \sqrt{{\vec P}^2 + M^2}} F_{X {\vec P} M^2}.
\ee We note, that in order to achieve a numerically 'flat'
function $\Gamma_X$ in space and time, one has to consider
averages over a large set of ensembles typically in the order of
$10^3$ testparticles per nucleon. By storing $\Gamma_X$ on a
4-dimensional grid the space-time derivatives of $\Gamma_X$, that
enter the equations of motion (\ref{eomrf}) - (\ref{eomm1f}), can
be evaluated in first or second order. We mention that the
Gaussian smearing algorithm described in Ref. \cite{CMMN} leads to
sufficiently stable results (see below). \\
The collisions of nucleons are described by the closest distance
criterion of Kodama et al. \cite{Kodama} in the individual $NN$
c.m.s., i.e.
\be
|{\vec X}_1 - {\vec X}_2| \leq \sqrt{\sigma({\vec P}_1 - {\vec P}_2)/\pi}
\ee
using the Cugnon parametrizations \cite{Cugnon} for the in-medium
$NN$ cross section $d\sigma/d\Omega(\sqrt{s})$ and identifying
(in the $NN$ c.m.s.)
\be
s- 4 m_N^2 = s-4M^2 = 4 {\vec P}^2,
\ee
where $m_N$ is the nucleon vacuum mass, $M$ the actual off-shell
mass and $\sqrt{s}$ the invariant energy of a nucleon-nucleon
collision in the vacuum with c.m.s. momentum ${\vec P}$. \\
According to eq. (\ref{Icoll}) the nucleons can change their virtual
mass $M$ in the scattering process $1 + 2 \rightarrow 3 + 4$, while
keeping the energy and momentum balance. This process is technically
handeled by selecting the final nucleon masses by Monte Carlo according
to the local Breit-Wigner distribution. However, our Monte Carlo
simulations showed that this change of virtuality for elastic collisions
has a minor effect on the observables to be discussed below. \\
Apart from the description of particle propagation and
rescattering the results of the transport approach also depend on
the initial conditions, ${\vec X}_i(0), {\vec P}_i(0), M_i^2(0)$.
In view of nucleus-nucleus collisions, i.e. two nuclei impinging
towards each other with a laboratory momentum per particle
$P_{lab}/A$, the nuclei can be considered as in their respective
groundstate, which in the semiclassical limit is given by the
local Thomas-Fermi distribution \cite{CMMN}. Additionally the
virtual mass $M_i^2$ has been determined by Monte-Carlo according
to the Breit-Wigner distribution (\ref{nonrelbreitwigner})
assuming an in-medium width $\Gamma_0 $ = 1 MeV. For the vacuum
width of the nucleons we have used $\Gamma_V $ = 1 MeV which
implies that nucleons propagating to the continuum in the final
state of the reaction achieve their vacuum mass on the 0.1 $\%$
level. We note in passing that for the initialization we
additionally have required $P_{0i} < m_N$ (in the restframe of the
nucleus) which due to energy conservation implies that particles
cannot escape 'numerically' from the nucleus in the groundstate. \\
\subsection{Nucleus-nucleus collisions at GANIL energies}
Our first applications we devote to nuclear reactions at GANIL
(92 - 95 A MeV) since here the more recent measurements have lead
to conflicting results between different transport approaches
\cite{Germain}. We start with the reaction $Ar + Ta$ at 92 A MeV. \\
In view of Section 3 we present for some randomly chosen testparticles
$i$ their off-mass-shell behaviour $M_i^2(t)-M_0^2$ as a function
of time in a central collision ($b$ = 1 fm) in Fig. 7. It is seen
that during the maximum overlap of the nuclei at $t \approx$ 30 fm/c
the off-shellness reaches up to 0.2 GeV$^2$, however, in analogy
to the model studies in Section 3 the nucleons become practically
on-shell for $t \ge$ 90 fm/c. The finite width at the end of the
calculation presented here is due to the fact that the collisional
width $\Gamma_{coll}$ is still different from zero. The fluctuations
in $M_i^2(t)-M_0^2$ in time give some idea about the numerical accuracy
of the calculation for the space-time derivative of $\Gamma_X$; the
functions become smoother when increasing the number of
testparticles/nucleon furtheron $(\geq$ 1000). \\
Without explicit representation we note that the proton rapidity
spectra $dN/dy$ do not change within the numerical accuracy when
comparing the on-shell propagation limit with the results from the
off-shell transport approach for $Ar + Ta$ at 92 A MeV. There is,
however, a small enhancement in the proton transverse momentum
spectra $1/p_T \, dN/dp_T$ for the off-shell propagation of
nucleons as can be seen in Fig. 8, when the proton $p_T$ spectra
are compared at an impact parameter $b$ = 1 fm. \\
The question remains, if such an enhancement might be seen experimentally
or if the off-shell approach overpredicts the high momentum tail of the
spectra. For this purpose we compare to the proton spectra from Ref.
\cite{exp} taken at $\theta_{lab}$ = 75$^o$ for kinetic energies above
175 MeV (Fig. 9). For this comparison we have integrated the proton spectra
over all impact parameters with stepsize $\Delta b$ = 1 fm in the angular
range $70^o \leq \theta_{lab} \leq 80^0$. The calculated spectra (open
squares in Fig. 9) only extend up to 200 MeV due to statistics, however,
match with the experimental data within the errorbars. A similar comparison
has been performed by Germain et al. \cite{Germain} where the traditional
BUU and QMD calculations seem to be compatible with the experimental data
within the statistics achieved whereas the Boltzmann Langevin (BL)
calculations incorporating fluctuations in momentum space \cite{BL}
overestimate the proton spectra by more than three orders of
magnitude \cite{Germain}. This rules out the latter BL approach but does
not imply that our off-shell transport approach will properly describe the
experimental spectra up to $E_{kin}$ = 350 MeV. In view of Fig. 9 the
calculational statistics would have to be increased by more than 3 orders
of magnitude which is unlikely to achieve for our present off-shell approach
within a reasonable time due to the high amount of processor capacity needed.
Note that due to nonlocal effects in the collision term the proton spectra
might be slightly hardened additionally \cite{Mora3}. \\
We thus continue with more qualitative investigations, that allow to
extract the physics more clearly. In Fig. 10 we display the number of
baryon-baryon collisions $dN^{BB}/d\sqrt{s}$ as a function of the
invariant energy $\sqrt{s}$ for $Ar + Ta$ at 92 A MeV integrated over
all impact parameters. The dashed line shows the result for the on-shell
transport approach (starting at $2 m_N$) whereas the solid line
corresponds to the off-shell result, which extends down to $\sqrt{s}
\approx$ 1.5 GeV. Note that elastic collisions of off-shell nucleons
can occur due to their dynamical virtuality in mass. The dashed line
is practically identical to the BUU calculations from Ref. \cite{Germain}
and is limited to collisions far below the kaon production threshold
of $\sqrt{s_0}\approx$ 2.54 GeV. Thus the kaon production yield of
(2.9$ \pm 1.6)\cdot 10^{-9} b$ claimed in Ref. \cite{Lecolley} for
this system cannot be described in the on-shell limit, however, also
not in our off-shell approach which shows only a small enhancement
in the high $\sqrt{s}$ regime. \\
The latter $\sqrt{s}$ distribution can approximately be tested
experimentally by hard photon spectra, a question that has been
explored by the TAPS collaboration for $Ar + Au$ at 95 A MeV
\cite{TAPS,Holzmann}. In order to test our transport approach
we have performed calculations for this system, too, using the
parametrizations (4.13) of Ref. \cite{CMMN} for the elementary
differential photon cross section in proton-neutron ($pn$) collisions.
Note that the elementary photon bremsstrahlung in $pn$ collisions
is at best known within a factor of 2 (cf. the discussion in Ref.
\cite{CMMN}). Fig. 11 displays the results of our bremsstrahlung
calculations in comparison to the data from Ref. \cite{TAPS}.
The dashed line corresponds to the conventional on-shell
calculation and is practically identical to the BUU analysis
performed by Holzmann et al. \cite{Holzmann}, but underestimates
the high energy photon yield dramatically. This situation does not
improve very much when including the off-shell propagation of
nucleons for the initial channel (dash-dotted line), however,
still requiring that the nucleons in the final state are on-shell,
too. Denoting off-shell nucleons by an extra  $^*$ this corresponds
to the individual reactions $p^*+ n^* \rightarrow p+n+\gamma$,
whereas the dashed line is obtained from the channel
$p+n \rightarrow p+n+\gamma$. \\
On the other hand, the energetic photons are produced very early in
the collision phase where the virtual mass distribution of nucleons
-- determined by $\Gamma_X$ (\ref{gcoll}) -- becomes very broad.
Thus including this virtuality in mass also in the final state,
where the masses are selected by Monte-Carlo according to
(\ref{nonrelbreitwigner}) with a local width $\Gamma_X$,
we selfconsistently can sum the individual channels $p^* + n^*
\rightarrow p'^* + n'^* + \gamma$. The result of such calculations
is shown in Fig. 11 by the solid line which comes quite close to
the experimental data \cite{TAPS}. We note that in the latter
calculations we have averaged the photon yield over 10 MeV bins to
reduce the statistical fluctuations emerging from the Monte-Carlo
final state selection. Whereas in Ref. \cite{Holzmann} the high
energy photon yield has been tentatively attributed to very high
momentum components in the initial phase-space distribution - which
semiclassically are not bound - our present results indicate that
this yield might be almost entirely explained (without introducing
any additional assumptions) by the off-shell transport approach.
It is presently unclear, if the missing high energy photon yield
should be attributed to three-body reaction channels
\cite{Bonasera,Gulmi}, to the contribution of the
$\Delta \rightarrow \gamma N$ channel \cite{Stachel} or to the
secondary $\pi N \rightarrow \gamma N$ channel \cite{Gudima,66a}. \\
The question now arises if the kaon yield from Ref. \cite{Lecolley}
for $Ar + Ta$ at 92 A MeV might also be due to off-shell hadronic
states in the final channel. We thus have performed calculations for
this system again within the following assumptions:
$\Gamma_X^N = \Gamma_X^{\Lambda}$ and $\Gamma_X^K$ = 0 since the $KN$
cross section is rather small \cite{C97}. The parametrisations for
the elementary reactions $NN \rightarrow N \Lambda  K$ and
$\pi N \rightarrow K \Lambda$ have been adopted from Refs.
\cite{C97,B97} where $K, \bar{K}$ production has been
systematically investigated for nucleus-nucleus reactions in the
SIS energy regime. In view of Fig. 11 the small enhancement of
energetic collisions for off-shell nucleons in the entrance
channel does not have a large effect on the kaon production cross
section as in case of energetic photons (cf. Fig. 11), but the
reduction of the threshold due to the virtuality in mass of the
final nucleon and $\Lambda$ hyperon should have (cf. Fig. 11).
Unfortunately, within the statistics achieved in our calculations
we did not find any $K \Lambda$ production event such that we
presently cannot provide a solid answer. \\
In order to obtain some upper estimate we parametrize the collisional
distribution $dN^{pn}/d\sqrt{s}$ (cf. Fig. 10) for $pn$ collisions by
an exponential tail
\be
\label{tail}
\frac{dN^{pn}}{d\sqrt{s}} \sim \exp\{-\frac{\sqrt{s}}{E_0}\}
\ee
with some slope parameter $E_0$ and fix the normalization constant as
well as $E_0$ by the $\gamma$ spectra from the TAPS collaboration.
The result of this model study is shown in Fig. 12 where the computed
photon spectra for $E_0$ = 33 MeV (dashed line) and $E_0 =$ 35 MeV
(solid line) are shown in comparison to the data \cite{TAPS}. We note
that within this model the photon spectra remain roughly exponential
up to 700 MeV photon energy, which would have to be proven by
experiment explicitly. \\
Now replacing the elementary differential photon cross section $d
\sigma_{pn \rightarrow pn \gamma}(\sqrt{s})/d\Omega$ by the
elementary kaon production cross section $d\sigma_{NN \rightarrow
K\Lambda N}(\sqrt{s})$ \cite{C97} and correcting for isospin we
obtain an inclusive kaon production cross section $\sigma_K
\approx 1.6 \cdot 10^{-11} b$ ($E_0 = 33$ MeV)  and $\approx 6
\cdot 10^{-11} b$ ($E_0 = 35$ MeV) for $Ar + Ta$ at 92 A MeV,
which is about two orders of magnitude smaller than the cross
section from Ref. \cite{Lecolley}. Moreover, a note of caution has
to be added here: In view of the analysis in Refs.
\cite{CB99,Schaffner,C97,B97,Li97f,waas} the kaon potential in the
nuclear medium is most likely repulsive and the $\Lambda$
potential only 2/3 of the nucleon potential. This shift in
production threshold to higher $\sqrt{s}$ due to the real part of
the hadron self-energies makes kaon production even more unlikely.
We thus do not expect to describe the kaon cross section from Ref.
\cite{Lecolley} in our off-shell approach even when increasing the
statistics by some orders of magnitude. \\
\section{Summary}
In this work we have derived and developed a semiclassical
transport approach that in first order in the gradient expansion
describes the virtual propagation of particles in the invariant
mass squared $M^2$ besides the conventional propagation in the
mean-field potential given by the real part of the self-energy.
The derivation has been based on the familiar Kadanoff-Baym
equations \cite{kb62} by exploiting the relations between the
different Green functions and the relations between their real and
imaginary parts. Whereas in conventional transport approaches the
imaginary part of the self-energy is reformulated in terms of a
collision integral and simulated by on-shell binary collisions, we
additionally account for the off-shell propagation of particles
due to the imaginary part of the self-energy in eqs. (\ref{eomrf})
- (\ref{eomm1f}). We note that in our formulation the
single-particle energy $P_0$ is fixed by eq. (\ref{energyfix}) and
that  the propagation is determined by $dP_0/dt$ = 0 in the
collisionless limit (if $Re \Sigma_X^{ret}$ has no explicit time
dependence). On the other hand, the local collision rate
$\Gamma_X$ is determined by the collision integrals themselves and
can be used in transport approaches without introducing any new
assumptions or parameters. In our present approach we have
restricted to momentum-independent imaginary self-energies; an
extension to the general case appears straight-forward according
to Section 2, however, is numerically much more involved. \\
As a first application we have studied the dynamical evolution of
particles in a fixed complex potential -- having some similarities
to hadron-nucleus collisions without explicit collisions -- and
demonstrated the off-mass-shell propagation in a transparent way
for a variety of model cases. As also shown numerically, the
energy conservation strictly holds for the set of eqs.
(\ref{eomrf}) - (\ref{eomm1f}). Furthermore, a distribution of
off-shell particles regains its vacuum spectral function when
moving out of the complex scattering centre; particles with
vanishing (or very small) width become asymptotically on-shell
again as required by the quantum mechanical boundary conditions
while in the absorptive medium the spectral function reproduces
the correct width in line with quantum mechanics (cf. Fig. 6). \\
We have, furthermore, presented the first dynamical calculations of
the novel transport theory for nucleus-nucleus collisions at GANIL
energies where we can test its results in comparison to experimental
data. We find that the off-shell propagation of nucleons practically
does not change the rapidity distributions $dN/dy$ and only has a
moderate effect on the high transverse momentum spectra of protons.
The latter we found to be fully compatible with the data from Ref.
\cite{exp} for $Ar + Ta$ at 92 A MeV contrary to the Boltzmann-Langevin
(BL) calculations in Ref. \cite{Germain}. The distribution of
nucleon-nucleon collisions in the invariant energy $\sqrt{s}$ is
found to be also slightly enhanced for high invariant energies
(as well as below the 2 nucleon threshold), which has some effect
on the production of high energy $\gamma$-rays. Here we controlled
our calculations by the photon data from the TAPS collaboration for
$Ar + Au$ at 95 A MeV \cite{TAPS} that could be reasonably described
in the off-shell limit when including especially the nucleon spectral
functions in the exit channel $p^* + n^* + \gamma$. We have argued
that this might be a first experimental indication for the off-shell
propagation of nucleons that has to occur due to quantum mechanics.
Our attempt to calculate the kaon cross section for $Ar + Ta$ at
92 A MeV \cite{Lecolley} within the same line  failed due to the
limited statistics. However, when extrapolating the collisional
$\sqrt{s}$ distribution by an exponentail tail (\ref{tail}) and
fixing the slope parameter by the photon data from the TAPS
collaboration \cite{TAPS} our upper limit for this cross section
is roughly two orders of magnitude below the experimental value,
which implies that the data point \cite{Lecolley} remains
ununderstood theoretically furtheron. \\
The actual applications of the present off-shell transport approach
are not limited to nuclear physics problems, but should be of relevance
for any system including particles of finite life time and/or high
collision rates. The more practical point is now to set up new
numerical recipies to increase statistics and also to include the
explicit momentum dependence stemming from the imaginary part of
the particle self-energy. \\
\vspace{1cm} \\
The authors like to acknowledge stimulating discussions with C. Greiner
and S. Leupold throughout this study. Furthermore, they like to thank
R. Holzmann for providing the data from Ref. \cite{TAPS} in electronic
form and for valuable explanations of the TAPS experiment.
%

%
%
\begin{figure}[h]
{\vspace*{-1.0 cm} \hspace*{0.0 cm}
{\epsfig{file=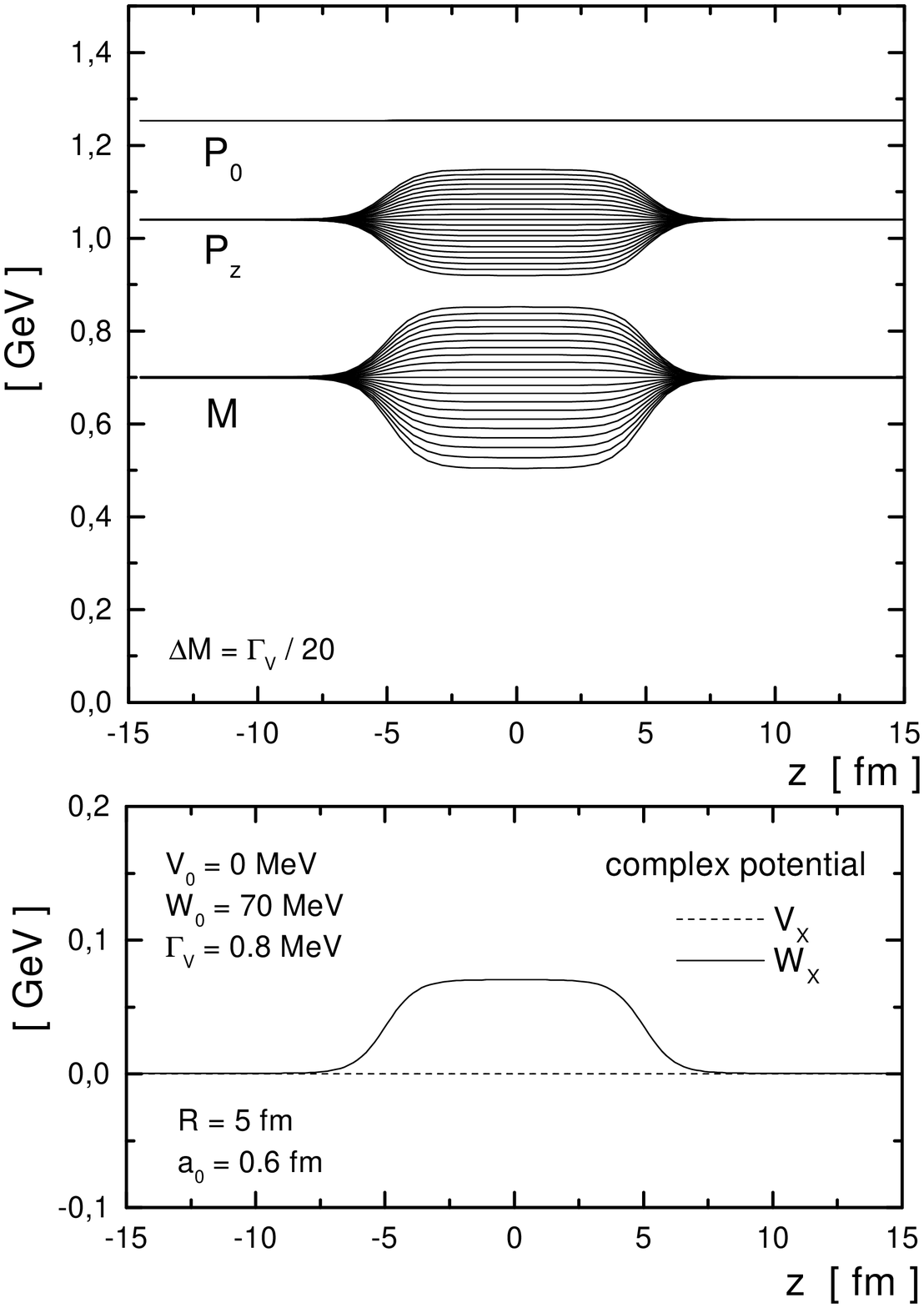,height=20 cm,width=15 cm}}} \caption{upper
part: $P_{i0}$, $M_{i}$ and $P_{iz}$ as a function of $z(t)$ for a
purely imaginary potential $W_{0}(t) = W_{0} = 70$ MeV (lower
part). The vacuum width is $\Gamma_{V} = 0.8$ MeV and the initial
separation in mass of the testparticles is given by $\Delta M =
0.05 \cdot \Gamma_{V}$. \label{bild1}}
\end{figure}
\begin{figure}[h]
{\vspace*{-1.0 cm} \hspace*{0.0 cm}
{\epsfig{file=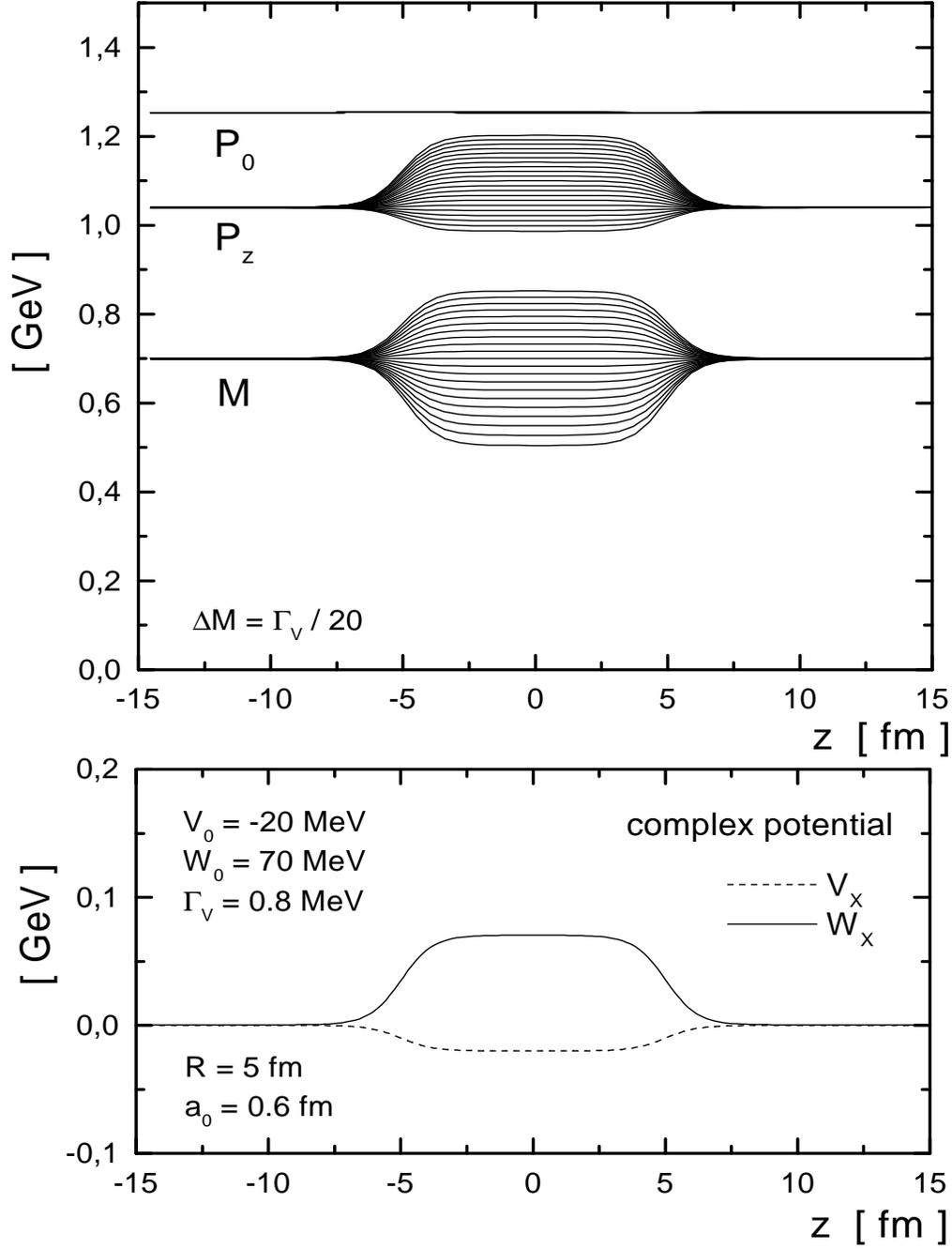,height=20 cm,width=15 cm}}} \caption{upper
part: $P_{i0}$, $M_{i}$ and $P_{iz}$ as a function of $z(t)$ for a
complex potential with  $V_{0} = -20$ MeV, $W_{0}(t) = W_{0} = 70$
MeV (lower part). For the vacuum width and the initial mass
separation we have used the same values as in Fig. 1.
\label{bild2}}
\end{figure}
\begin{figure}[h]
{\vspace*{-1.0 cm} \hspace*{0.0 cm}
{\epsfig{file=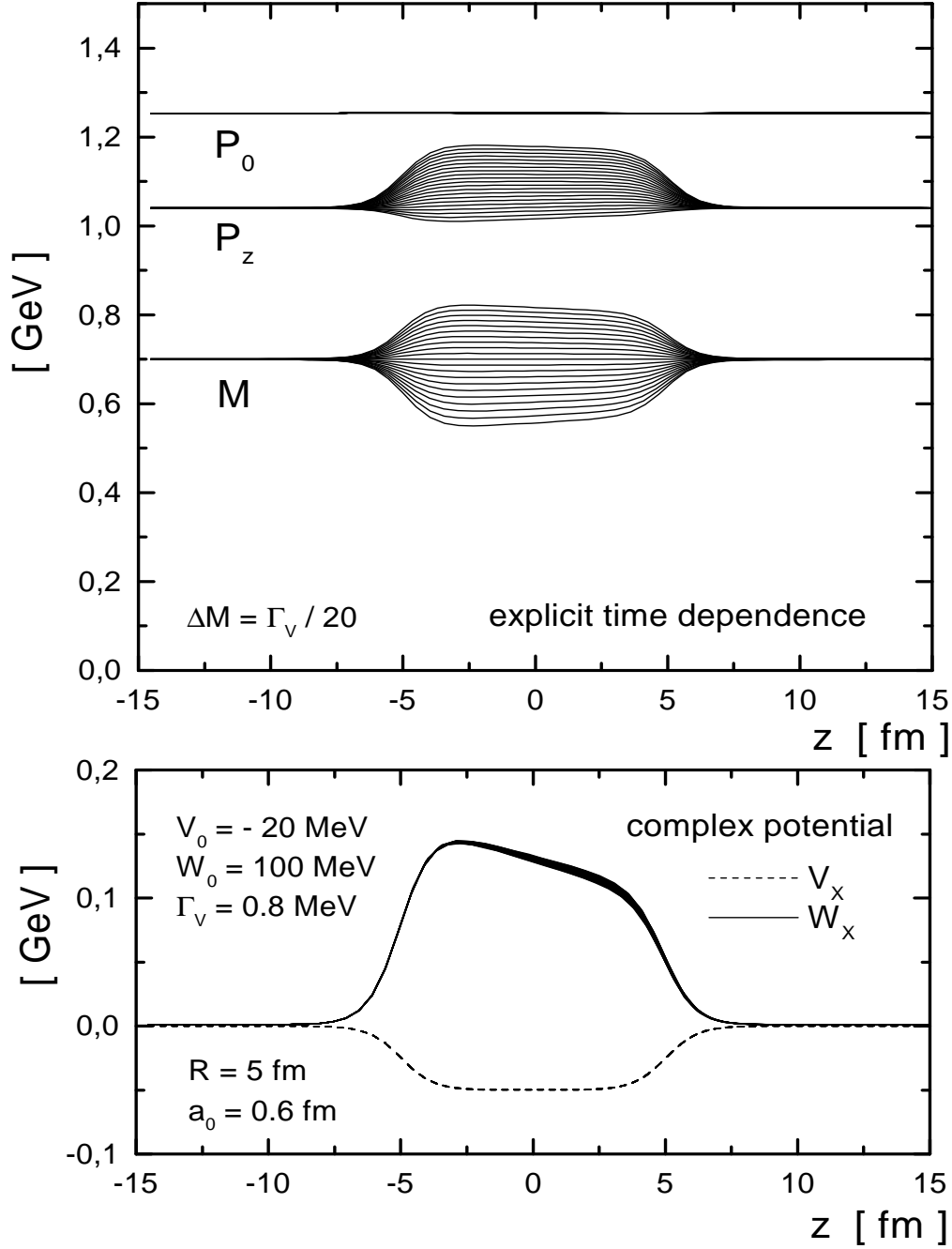,height=20 cm,width=15 cm}}}
\caption{upper part: $P_{i0}$, $M_{i}$ and $P_{iz}$ as a function
of $z(t)$ for an explicitly time-dependent imaginary part of the
potential $W_{0}(t) = 100$ MeV $(1-0.02 \, c/$fm$ \cdot t)$.
The real part is taken as in Fig. 2 (lower part).
For the vacuum width and the initial mass seperation we have used
the same values as in Fig. 1.
\label{bild3}}
\end{figure}
\begin{figure}[h]
{\vspace*{-1.0 cm} \hspace*{0.0 cm}
{\epsfig{file=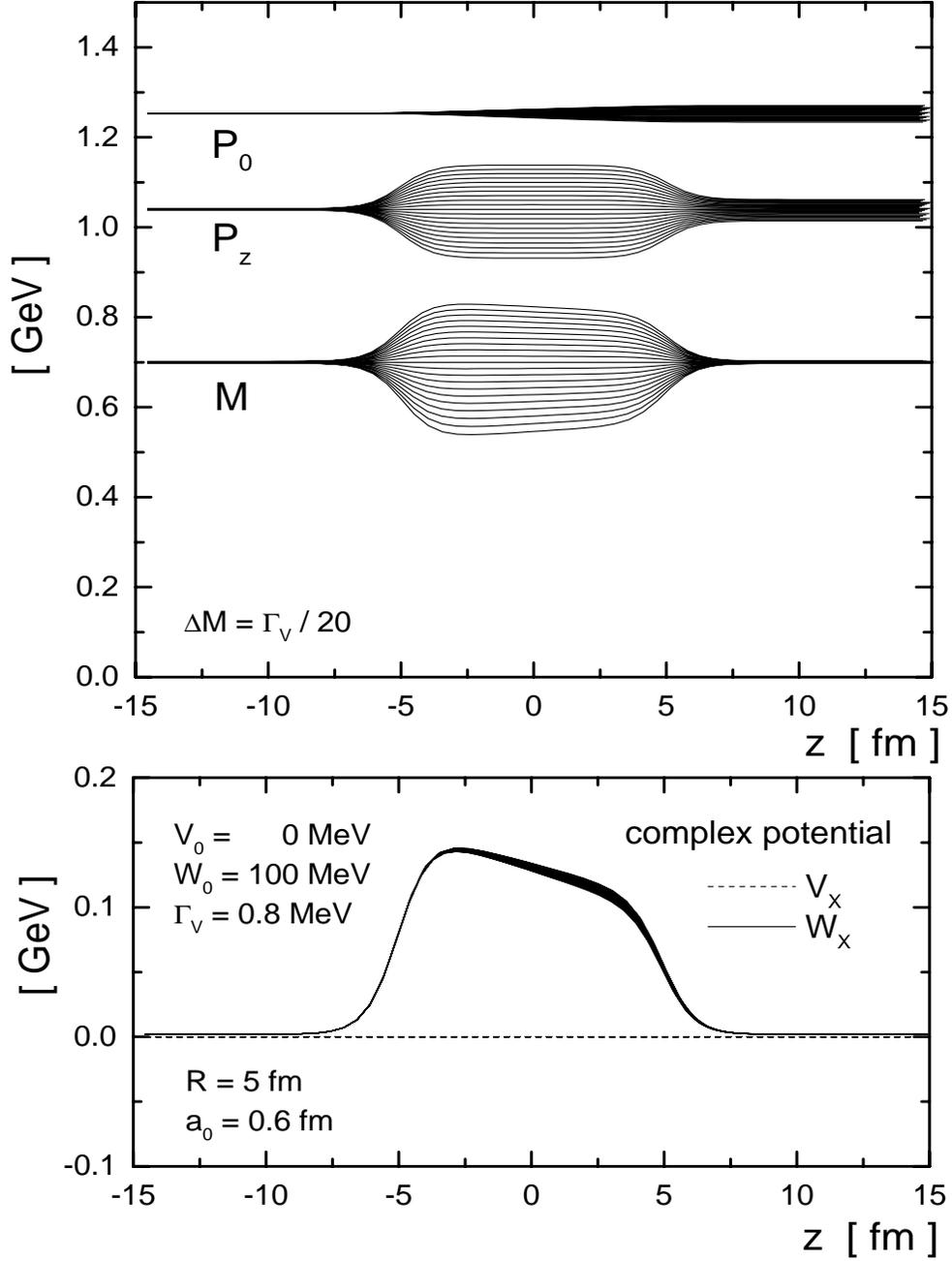,height=20 cm,width=15 cm}}} \caption{
$P_{i0}$, $M_{i}$ and $P_{iz}$ as a function of $z(t)$ for the
same time-dependent imaginary part of the potential $W_{0}(t)$ as
in Fig. 3 but discarding the term $\sim
\partial \Gamma_X /\partial t$ in eq. (\protect\ref{eompf}).
\label{bild3b}}
\end{figure}
\begin{figure}[h]
{\vspace*{-1.0 cm} \hspace*{0.0 cm}
{\epsfig{file=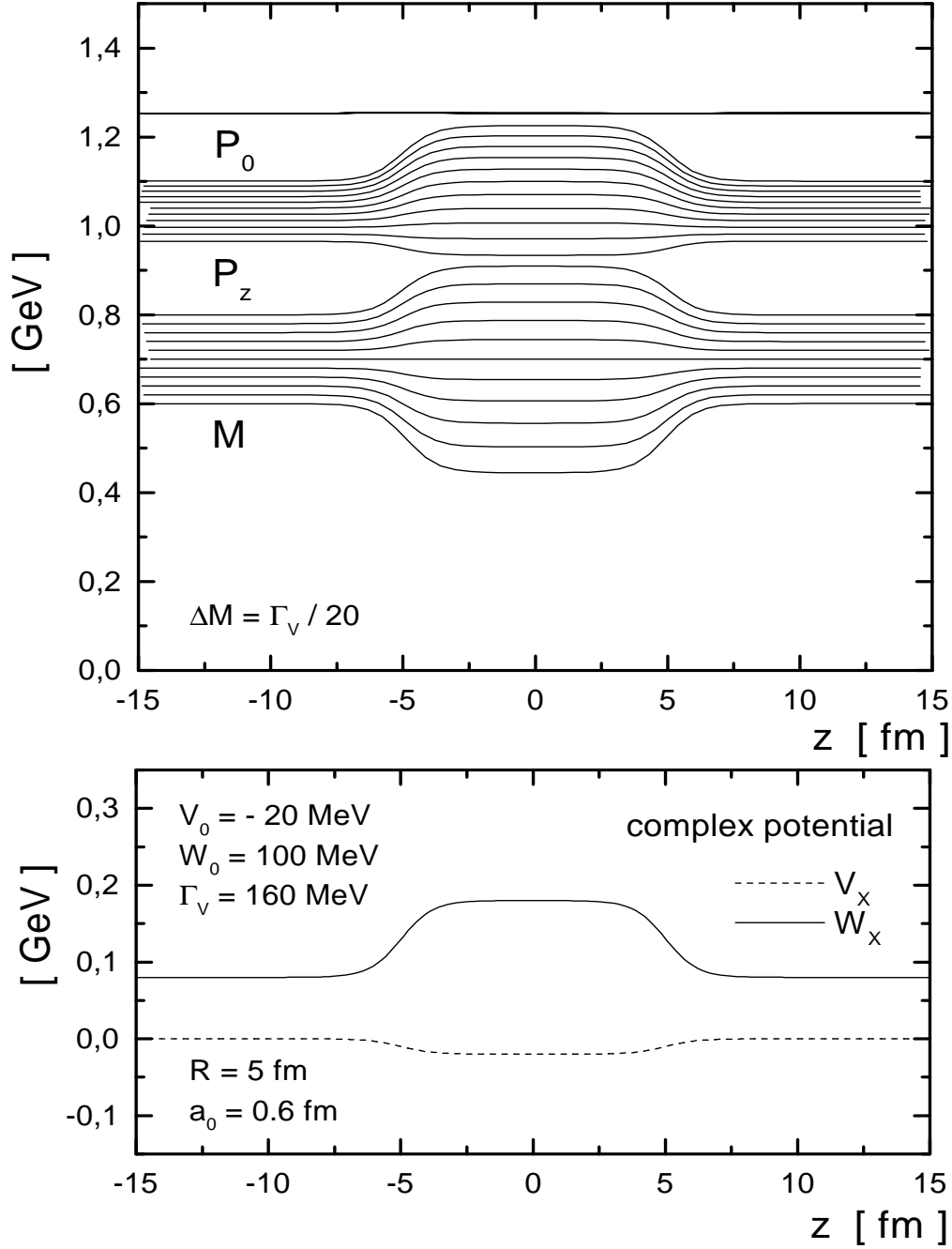,height=20 cm,width=15 cm}}}
\caption{upper part: $P_{i0}$, $M_{i}$ and $P_{iz}$ as a function of
$z(t)$ for a broad vacuum spectral function in a time-independent
potential $V_{0} = -20$ MeV, $W_{0}(t) = W_{0} = 100$ MeV
(lower part). We have chosen a vacuum width $\Gamma_{V} = 160$ MeV
and an initial mass seperation of $\Delta M = 0.05 \cdot
\Gamma_{V}$ for the testparticle trajectories displayed.
\label{bild4}}
\end{figure}
\begin{figure}[h]
\vspace*{-1.0 cm} \hspace*{0.0 cm}
\epsfig{file=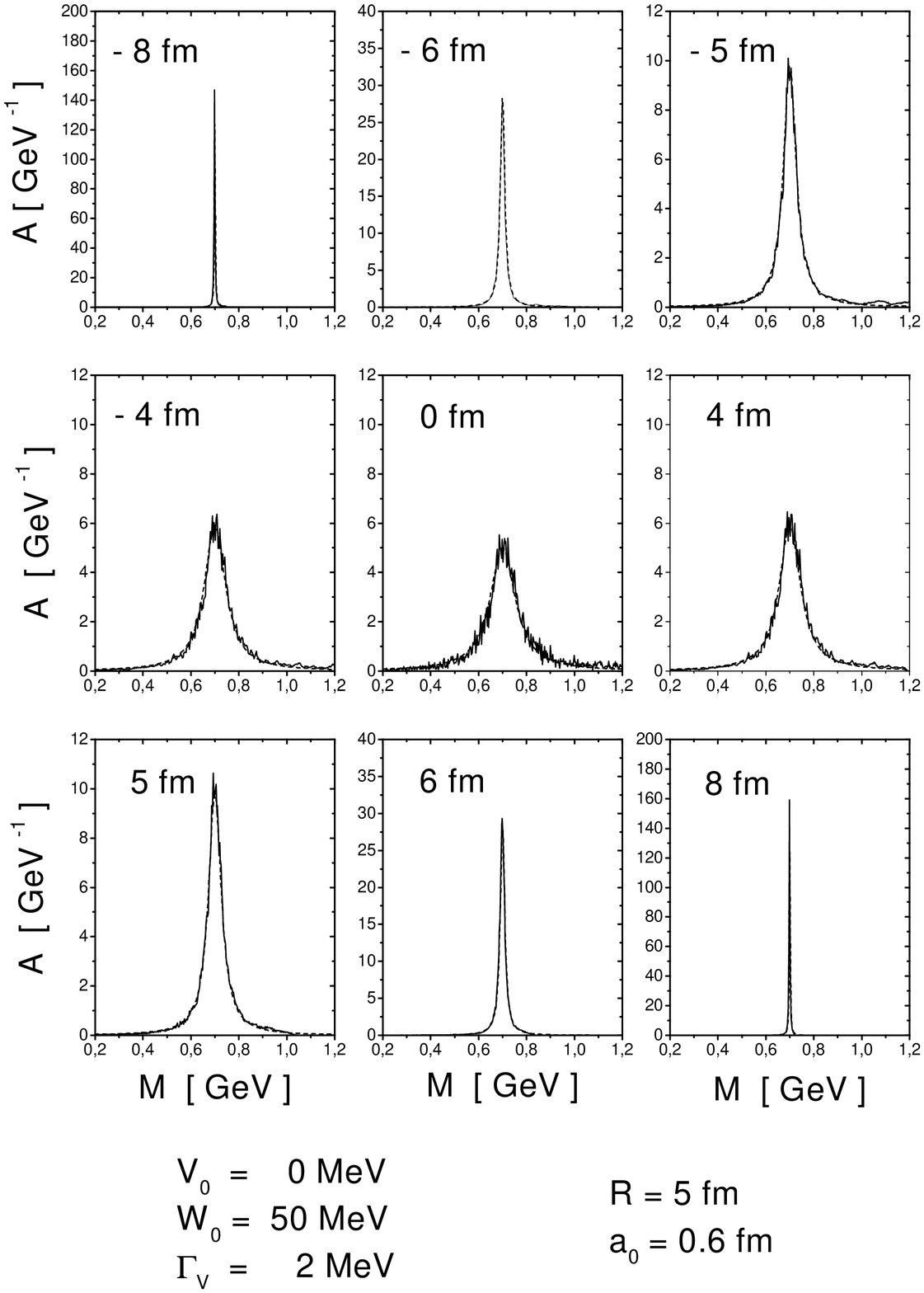,height=20cm,width=15cm}
\caption{The spectral distribution at different coordinates $z$
from the testparticle distribution in comparison to the analytical
result (solid lines) for $V_0$ = 0, $W_0$ = 50 MeV and $\Gamma_V$ =
2 MeV. The analytical result is practically identical to the histograms from
the testparticle distribution and thus hardly visible.}
\label{bild4b}
\end{figure}
\begin{figure}[h]
{\vspace*{-1.0 cm} \hspace*{0.0 cm}
{\epsfig{file=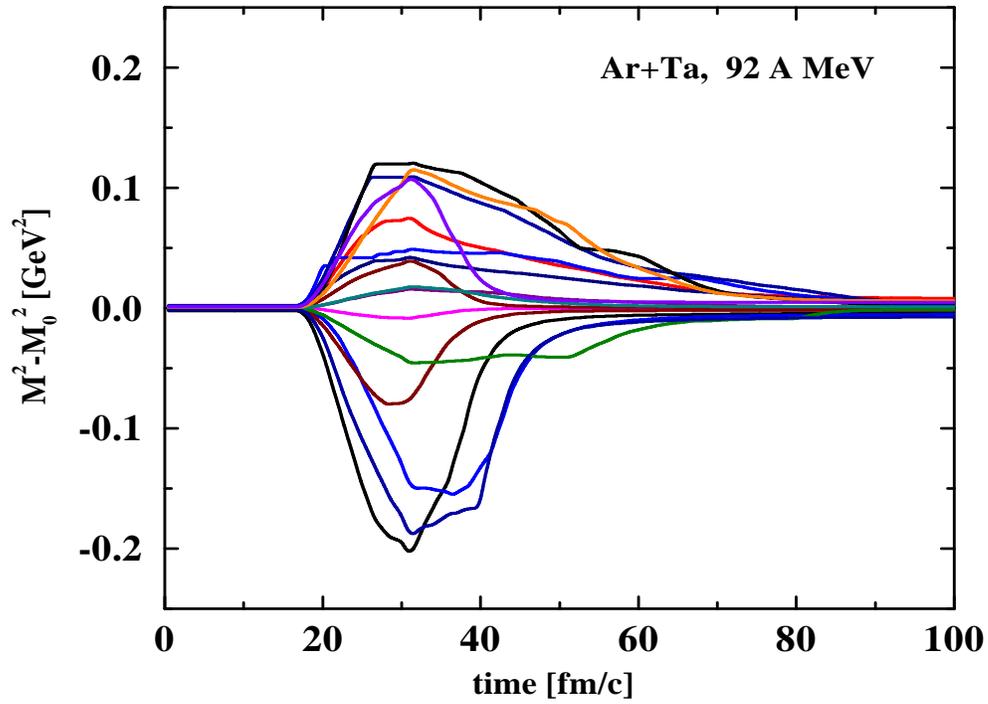,height=20 cm,width=15 cm}}}
\caption{The off-shell propagation in mass ($M_i^2(t)-M_0^2$) as a
function of time for 16 randomly chosen testparticles. The system
is $Ar + Ta$ at 92 A MeV and impact parameter $b$ = 1 fm.
\label{bild5}}
\end{figure}
\begin{figure}[h]
{\vspace*{-1.0 cm} \hspace*{0.0 cm}
{\epsfig{file=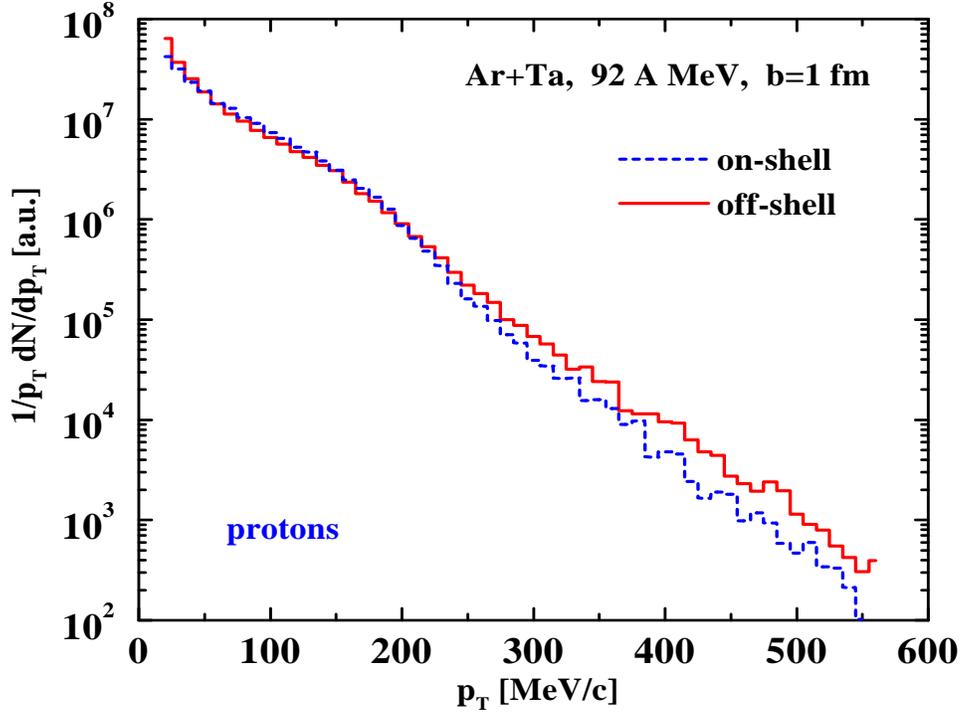,height=20 cm,width=15 cm}}}
\caption{The transverse momentum spectra of protons for $Ar + Ta$
at 92 A MeV and impact parameter $b$ = 1 fm. The dashed histogram
is the result from the on-shell propagation while the solid
histogram is obtained including the off-shell propagation of
nucleons. \label{bild6}}
\end{figure}
\begin{figure}[h]
{\vspace*{-1.0 cm} \hspace*{0.0 cm}
{\epsfig{file=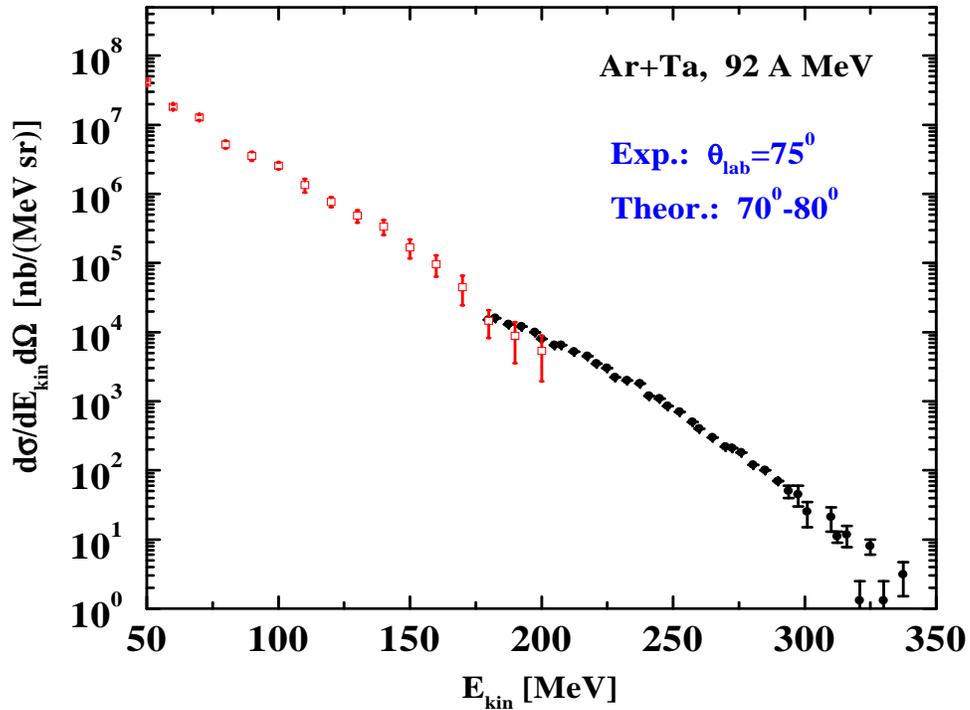,height=20 cm,width=15 cm}}}
\caption{The inclusive kinetic energy spectra of protons for $Ar +
Ta$ at 92 A MeV including the off-shell propagation in the
transport approach (open squares) for $70^0 \leq \theta_{lab} \leq
80^0$ in comparison to the experimental data from
\protect\cite{exp} (full circles). \label{bild7}}
\end{figure}
\begin{figure}[h]
{\vspace*{-1.0 cm} \hspace*{0.0 cm}
{\epsfig{file=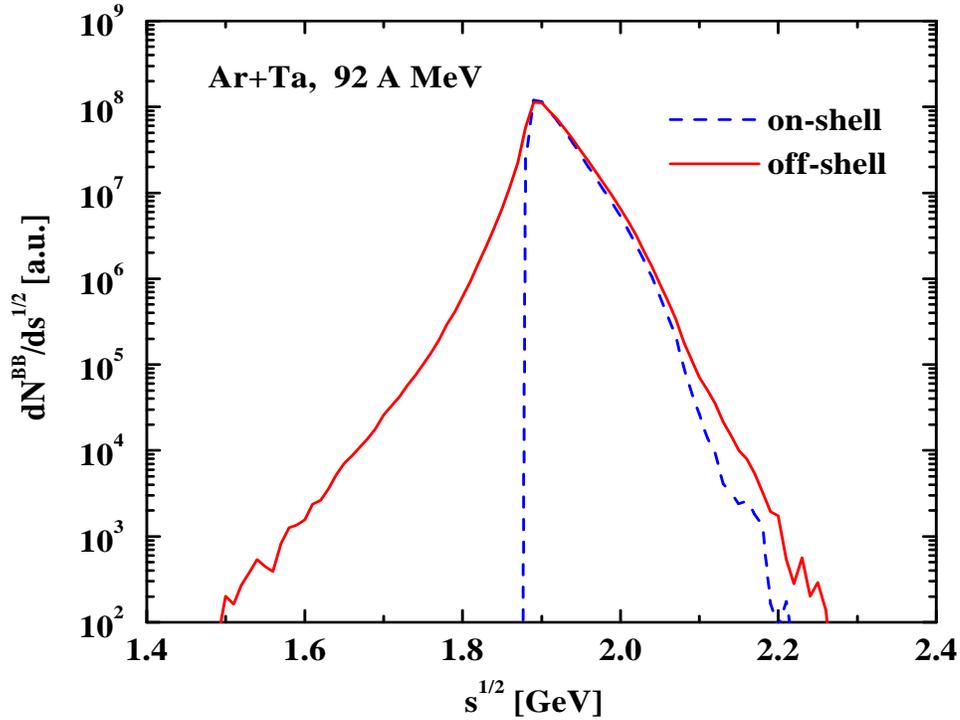,height=20 cm,width=15 cm}}}
\caption{The number of baryon-baryon (BB) collisions as a function of the
invariant energy $\sqrt{s}$
for $Ar + Ta$ at 92 A MeV integrated over all impact parameter. The solid
line is obtained from including the off-shell propagation in the
transport approach while the dashed line stands for the result in the
on-shell limit.
\label{bild8}}
\end{figure}
\begin{figure}[h]
{\vspace*{-1.0 cm} \hspace*{0.0 cm}
{\epsfig{file=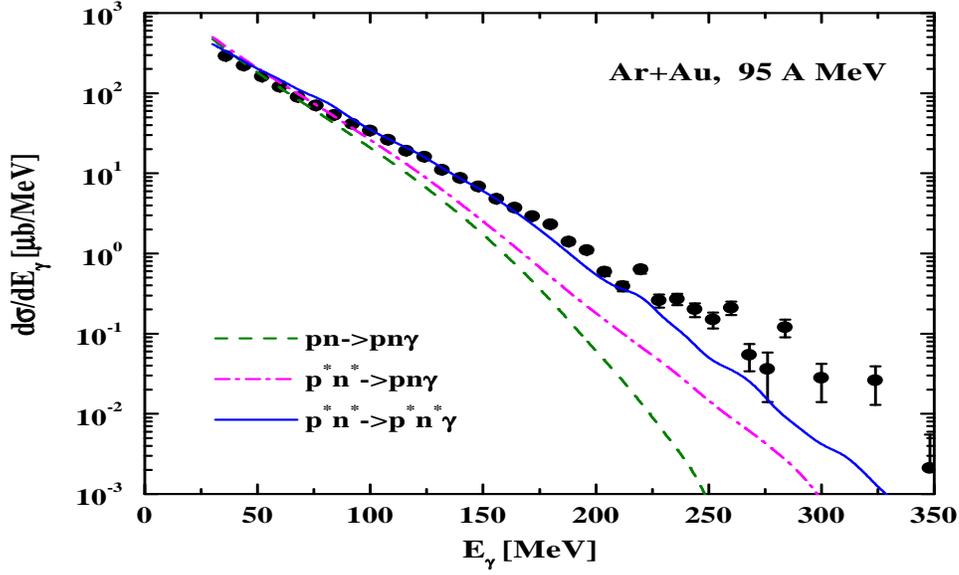,height=16 cm,width=15 cm}}}
\caption{The inclusive differential photon spectra
for $Ar + Au$ at 95 A MeV within various limits in comparison to the
data from Ref. \protect\cite{TAPS}. The dashed line is obtained in the
on-shell propagation limit including on-shell nucleons in the final state, too.
The dash-dotted line results from the off-shell propagation, however, including
on-shell nucleons in the final production channel. The solid line results from
the off-shell propagation of nucleons including also off-shell nucleons in
the final channel.
\label{bild9}}
\end{figure}
\begin{figure}[h]
{\vspace*{-1.0 cm} \hspace*{0.0 cm}
{\epsfig{file=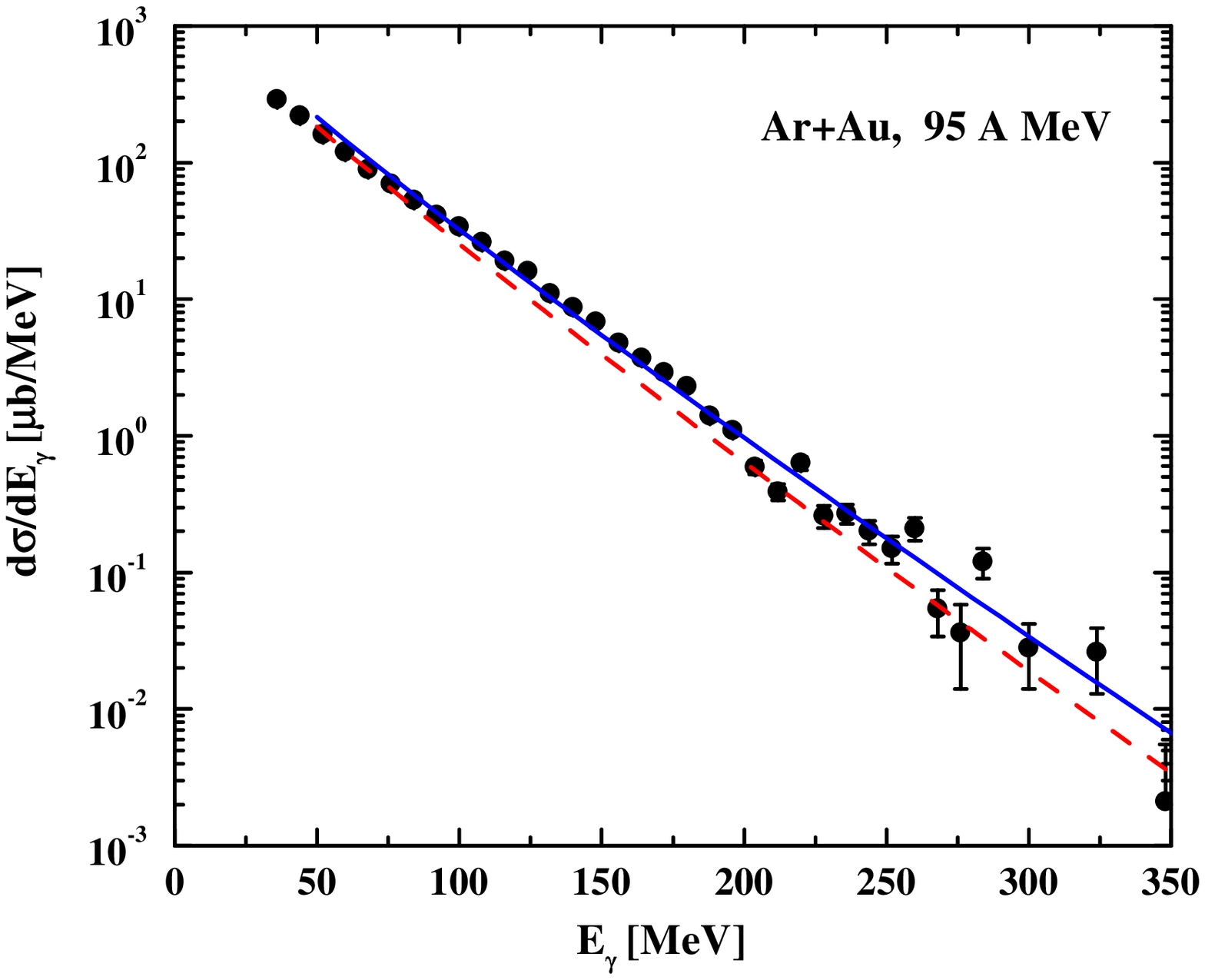,height=20 cm,width=15 cm}}}
\caption{The inclusive differential photon spectra
for $Ar + Au$ at 95 A MeV using the exponential $\sqrt{s}$ distribution
(\protect\ref{tail}) for $E_0$ = 33 MeV (dashed line) and $E_0$ = 35 MeV
(solid line) in comparison to the
data from Ref. \protect\cite{TAPS}.
\label{bild10}}
\end{figure}

\begin{thebibliography}{999}
\bibitem{js61}
   J. Schwinger, J. Math. Phys. {2} (1961) 407.
\bibitem{Wang1}
  S. J. Wang and W. Cassing, Ann. Phys. (N.Y.) 159 (1985) 328.
\bibitem{CaWa}
   W. Cassing and S. J. Wang, Z. Phys. A 337 (1990) 1.
\bibitem{cs85}
   K. Chou, Z. Su, B. Hao, and L. Yu, Phys. Rep. {118} (1985) 1.
\bibitem{md90}
   S. Mr\'{o}wczy\'{n}ski and P. Danielewicz, Nucl. Phys. B {342} (1990) 345.
\bibitem{mh94}
   S. Mr\'{o}wczy\'{n}ski and U. Heinz, Ann. Phys. (N.Y.) {229} (1994) 1.
\bibitem{Stoecker}
    H. St\"ocker and W. Greiner, Phys. Rep. 137 (1986) 277.
\bibitem{Bertsch}
    G. F. Bertsch and S. Das Gupta, Phys. Rep. 160 (1988) 189.
\bibitem{CMMN}
    W. Cassing, V. Metag, U. Mosel, and K. Niita,
    Phys. Rep. 188 (1990) 363.
\bibitem{Cass}
    W. Cassing and U. Mosel, Prog. Part. Nucl. Phys.  25 (1990) 235.
\bibitem{T3}
    A. Faessler, Prog. Part. Nucl. Phys. 30 (1993) 229.
\bibitem{URQMD}
    S. Bass, M. Belkacem, M. Bleicher et al.,
    Prog. Part. Nucl. Phys. 41 (1998) 255.
\bibitem{CB99}
    W. Cassing and E. L. Bratkovskaya, Phys. Rep. 308 (1999) 65.
\bibitem{kb62} L. P. Kadanoff and G. Baym,
    {\it Quantum statistical mechanics}, Benjamin, New York, 1962.
\bibitem{pd841}
   P. Danielewicz, Ann. Phys. (N.Y.) 152 (1984) 239; {\it ibid.} 305.
\bibitem{Bot}
    W. Botermans and R. Malfliet, Phys. Rep. 198 (1990) 115.
\bibitem{Mal}
    R. Malfliet, Prog. Part. Nucl. Phys. 21 (1988) 207.
\bibitem{ph95}
    P. A. Henning, Nucl. Phys. A 582 (1995) 633; Phys. Rep. 253 (1995) 235.
\bibitem{gl98}
    C. Greiner and S. Leupold, Ann. Phys. (N.Y.) 270 (1998) 328.
\bibitem{Zuo}
    S. J. Wang, W. Zuo and W. Cassing, Nucl. Phys. A 573 (1994) 245.
\bibitem{CNW}
    W. Cassing, K. Niita and S. J. Wang, Z. Phys. A 331 (1988) 439.
\bibitem{Fauser}
    R. Fauser and H. Wolter, Nucl. Phys. A 584 (1995) 604.
\bibitem{Koreview}
    C. M. Ko and G. Q. Li, J. Phys. G: Nucl. Part. Phys. 22 (1996) 1673.
\bibitem{Aich}
    J. Aichelin, Phys. Rep. 202 (1991) 233.
\bibitem{T2}
    G. Q. Li, A. Faessler, and S. W. Huang,
    Prog. Part. Nucl. Phys. 30 (1993) 159.
\bibitem{LV}
    C. Gr\'{e}goire, B. Remaud, F. Sebille, L. Vinet, and Y. Raffray,
    Nucl. Phys. A 465 (1987) 317
\bibitem{RQMD}
    H. Sorge, H. St\"ocker, and W. Greiner, Ann. Phys. 192 (1989) 266.
\bibitem{ART}
    B. A Li and C. M. Ko, Phys. Rev. C 52 (1995) 2037.
\bibitem{Kah2}
    S. H. Kahana, D. E. Kahana, Y. Pang, and T. J. Schlagel,
    Annu. Rev. Nucl. Part. Sci. 46 (1996) 31.
\bibitem{Ehehalt}
    W. Ehehalt and W. Cassing, Nucl. Phys. A 602 (1996) 449.
\bibitem{Ayik}
    S. Ayik and C. Gr\'{e}goire, Nucl. Phys. A 513 (1990) 187.
\bibitem{Randrup}
    J. Randrup and B. Remaud, Nucl. Phys. A 514 (1990) 339.
\bibitem{Burgio}
    G. F. Burgio and Ph. Chomaz, Nucl. Phys. A 529 (1991) 157.
\bibitem{BL}
    E. Suraud, S. Ayik, M. Belkacem and J. Stryjewski,
    Nucl. Phys. A 542 (1992) 141; \\
    M. Belkacem, E. Suraud and S. Ayik,
    Phys. Rev. C 47 (1993) R16; \\
    Y. Abe, S. Ayik, P.-G. Reinhard, and E. Suraud,
    Phys. Rep. 275 (1996) 49.
\bibitem{Rudy1}
    R. Malfliet, Nucl. Phys. A 545 (1992) 3.
\bibitem{Rudy2}
    R. Malfliet, Phys. Rev. B 57 (1998) R11027.
\bibitem{Pavel}
    P. Danielewicz and S. Pratt, Phys. Rev. C 53 (1996) 249.
\bibitem{Mora1}
    V. Spicka, P. Lipavsky and K. Morawetz, Phys. Rev. B 55
    (1997) 5095; Phys. Lett. A 240 (1998) 160.
\bibitem{Mora2}
    P. Lipavsky, V. Spicka and K. Morawetz, Phys. Rev. E 59 (1999) 1291.
\bibitem{Mora3}
    K. Morawetz, V. Spicka, P. Lipavsky, and Ch. Kuhrts, nucl-th/9902008.
\bibitem{Ehe93}
    W. Ehehalt, W. Cassing, A. Engel, U. Mosel, and Gy. Wolf,
    Phys. Rev. C 47 (1993) R2467.
\bibitem{Effe99}
    M. Effenberger, E. L. Bratkovskaya and U. Mosel, nucl-th/9903026.
\bibitem{Rapp}
    R. Rapp, G. Chanfray and J. Wambach, Nucl. Phys. A 617 (1997) 472.
\bibitem{Brown91}
    G. E. Brown and M. Rho, Phys. Rev. Lett. 66 (1991) 2720.
\bibitem{CaRa}
    W. Cassing, E. L. Bratkovskaya, R. Rapp and J. Wambach,
    Phys. Rev. C 57 (1998) 916.
\bibitem{exp}
    M. Germain et al., Nucl. Phys. A 620 (1997) 81.
\bibitem{TAPS}
    R. Holzmann et al., Phys. Rev. Lett. 72 (1994) 1608.
\bibitem{Lecolley}
    F. R. Lecolley et al., Nucl. Phys. A 583 (1995) 379c.
\bibitem{lk64}
    L. V. Keldysh, Zh. Eksper. Teoret. Fiz. {47} (1964) 1515;
    Sov. Phys. JETP {20} (1965) 1018.
\bibitem{Serot}
    B. D. Serot and J. D. Walecka, Adv. Nucl. Phys. 16 (1986) 1.
\bibitem{Kodama}
    T. Kodama, S.B. Duarte, K.C. Chung, R. Donangelo, R.A.M.S. Nazareth,
    Phys. Rev. C 29 (1984) 2146.
\bibitem{Cugnon}
     J. Cugnon, D. Kinet and J. Vandermeulen, Nucl. Phys. A 379 (1982) 553.
\bibitem{Germain}
    M. Germain, Ch. Hartnack, J. L. Laville, J. Aichelin, M. Belkacem, and
    E. Suraud, Phys. Lett. B 437 (1998) 19.
\bibitem{Holzmann}
    R. Holzmann et al., Proc. of the 7th International Conference on Nuclear
    Reaction Mechanisms, Varenna, June 6-11, 1994, ed. by E. Gadioli, p. 261.
\bibitem{Bonasera}
    A. Bonasera, F. Gulminelli and J. Molitoris, Phys. Rep. 243 (1994) 1.
\bibitem{Gulmi}
    A. Bonasera and F. Gulminelli, Phys. Lett. B 259 (1991) 399;
    B 275 (1992) 24.
\bibitem{Stachel} M. Prakash, P. Braun-Munzinger, J. Stachel and
N. Alamanos, Phys. Rev. C37 (1988) 1959.
\bibitem{Gudima}
    K. K. Gudima et al.,
    Phys. Rev. Lett. 76 (1996) 2412.
\bibitem{66a} G. Martinez et al., Preprint SUBATECH 1999.
\bibitem{C97}
    W. Cassing, E. L. Bratkovskaya, U. Mosel, S. Teis,
    and A. Sibirtsev, Nucl. Phys. A 614 (1997) 415.
\bibitem{B97}
    E. L. Bratkovskaya, W. Cassing and U. Mosel,
    Nucl. Phys.  A 622 (1997) 593.
\bibitem{Schaffner}
    J. Schaffner-Bielich, I. N. Mishustin and J. Bondorf,
    Nucl. Phys. A 625 (1997) 325.
\bibitem{Li97f}
    G. Q. Li, C.-H. Lee and G. E. Brown, Nucl. Phys. A 625 (1997) 372.
\bibitem{waas}
    T. Waas, N. Kaiser, and W. Weise, Phys. Lett.  B 379 (1996) 34.
\end{thebibliography}
\end{document}